\documentclass[preprint]{iucr}
\journalcode{A}  
\usepackage{graphicx} 
\usepackage{epsfig} 
\usepackage{dcolumn} 
\usepackage{bm} 
\usepackage{amsmath}
\usepackage{xcolor}
\usepackage{xspace}
\usepackage{textcomp}
\usepackage[normalem]{ulem}
\graphicspath{{./}{Figures//}}

\definecolor{robgreen}{HTML}{00a832}
\definecolor{nryellow}{HTML}{8a8500}
\definecolor{cpblue}{HTML}{2e665f}
\definecolor{robadd}{HTML}{005c12}
\definecolor{emiladd}{HTML}{1438c7}
\definecolor{black}{HTML}{000000}

\def\firstrevadd#1{\textcolor{black}{#1}}

\def\rjkadd#1{\textcolor{black}{#1}}
\def\esbaddold#1{\textcolor{black}{#1}}
\def\oldrobadd#1{\textcolor{black}{#1}}

\newcommand{\RAA}{\AA$^{-1}$}
\def\cis{CuIr$_{2}$S$_{4}$}
\def\fd3m{Fd$\overline 3$m}

\newcommand{\ia}{\AA\ensuremath{^{-1}}\xspace}
\newcommand{\qmax}{\ensuremath{Q_{\mathrm{max}}}\xspace}

\newcommand{\soutoldold}[1]{}
\newcommand{\soutold}[1]{}
\newcommand{\soutoldoldold}[1]{}

\begin{document}
\title{On single crystal total scattering data reduction \rjkadd{and correction} protocols for analysis in direct space}

\author[a]{Robert~J.}{Koch}
\author[b]{Nikolaj}{Roth}
\author[a]{Yiu}{Liu}
\author[c]{Oleh}{Ivashko}
\author[c]{Ann-Christin}{Dippel}
\author[a]{Cedomir}{Petrovic}
\author[b]{Bo~B.}{Iversen}
\author[c]{Martin~v.}{Zimmermann}
\cauthor[a]{Emil~S.}{Bozin}{bozin@bnl.gov}

\aff[a]{Condensed Matter Physics and Materials Science Division, Brookhaven National Laboratory, Upton, NY 11973, \country{USA}}
\aff[b]{Center for Materials Crystallography, Department of Chemistry and iNANO, Aarhus University, DK-8000, Aarhus, \country{Denmark}}
\aff[c]{Deutsches Elektronen-Synchrotron DESY, 22607 Hamburg, \country{Germany}}

\maketitle                        

\begin{synopsis}
We explore data reduction and processing as it related to single crystal diffuse scattering and 3D-$\Delta$PDF experiments.
\end{synopsis}

%
%
\begin{abstract}
\rjkadd{We explore} data reduction \rjkadd{and correction} steps and \rjkadd{processed data} reproducibility in the emerging single crystal total scattering based technique of three-dimensional differential atomic pair distribution function (3D-$\Delta$PDF) analysis\rjkadd{\soutold{ are explored}}.
All steps from sample measurement to data-processing are outlined in detail\rjkadd{\soutold{ with}} using a \cis\ example  \rjkadd{crystal\soutold{system}} studied in a setup equipped with a high-energy x-ray beam and a \rjkadd{flat panel area detector\soutold{image plate detector}}.
\oldrobadd{Computational overhead as it pertains to data-sampling and the associated data processing steps is also discussed.}
Various aspects of the \rjkadd{\soutold{data}final 3D-$\Delta$PDF} reproducibility are explicitly tested by varying data-processing order and\firstrevadd{\soutoldoldold{varying which processing steps are included} included steps}, and by carrying out a \rjkadd{\soutold{sample-to-sample}crystal-to-crystal} data comparison.
We \rjkadd{\soutold{find that 3D-$\Delta$PDF is robust}identify situations in which the 3D-$\Delta$PDF is robust, and caution against a few particular cases which can lead to inconsistent 3D-$\Delta$PDFs}.
\firstrevadd{Although not all the approaches applied here-in will be valid across all systems, and a more in-depth analysis of some of the effects of the data processing steps may still needed, the methods collected here-in represent the start of a more systematic discussion about data processing and corrections in this field.}
\end{abstract}

\date{\today}
\maketitle

\section{Introduction}
\label{section:intro}
\subsection{Background}
\label{section:background}
\esbaddold{\soutoldold{Recently}Over the past several decades}, \esbaddold{total scattering based structural} studies revealing deviations between the true local atomic structure and that obtained by averaging over relatively long length scales have become more common~\cite{egami;b;utbp03}.
This increased interest has been spurred by a convergence of many factors, including the availability of high throughput \esbaddold{synchrotron-based} user facilities~\cite{schlachter_third-generation_1994,bilderback_review_2005} with improved detectors~\cite{chupas_rapid-acquisition_2003, broennimann_pilatus_2006,kraft_pilatus_2010}
software development~\cite{proffen_discus:_1997,qiu;jac04i,soper_partial_2005,farrow_pdffit2_2007,neder_diffuse_2008,tucker_rmcprofile:_2007,juhas;jac13,
yang_xpdfsuite:_2014,coelho_fast_2015,ashiotis_fast_2015,juhas_complex_2015,hammersley_fit2d:_2016,aoun_fullrmc_2016}, and\esbaddold{, most significantly,} an increased \esbaddold{\soutoldold{understanding} awareness of the materials research community} that\rjkadd{\soutold{ these}} local deviations often \esbaddold{\soutoldold{can play a critical role}represent an important ingredient} in \esbaddold{the observed \rjkadd{material \soutold{physical}} properties in diverse classes of functional} materials~\cite{egami_local_1991,billinge_local_1994, frandsen_widespread_2018,davenport_fragile_2019,wang_tetragonal_2020}.

The \esbaddold{analysis} technique \esbaddold{based on the} atomic pair distribution function\esbaddold{\soutoldold{analysis}}, resulting from a powder system \oldrobadd{(one-dimensional (1D)-}PDF) has often been the tool of choice for such local structure studies~\cite{bozin_local_2019,koch_room_2019, yang_two-orbital_2020}.
The \oldrobadd{1D-}PDF measurement is effectively a modified powder diffraction experiment, where the total \esbaddold{\soutoldold{diffraction}scattering} signal $I_{\textrm{tot}}$ is collected to large momentum transfer values, $Q$, after which the properly corrected \esbaddold{and background subtracted} $I_{\textrm{tot}}(Q)$ signal is sine Fourier transformed (FT) to obtain the \oldrobadd{1D-}PDF, $G(r)$, which is proportional to the probability of finding \esbaddold{\soutoldold{two atom pairs}a pair of atoms in a material} separated by a \esbaddold{{\it scalar}} distance $r$.
The nature of the \firstrevadd{\soutoldoldold{sample}orientational averaging} dictates that neither $I_{\textrm{tot}}(Q)$ nor $G(r)$ contain any directional information on atom-pair correlations~\cite{egami;b;utbp03}\firstrevadd{\soutoldoldold{as this information is lost by orientational averaging}}.
\firstrevadd{\soutoldoldold{Spatial information must be}This directional information is often} retrieved by means of structural modeling~\cite{farro;jpcm07,juhas_complex_2015}.

Recently the concepts behind the \oldrobadd{1D-}PDF \esbaddold{\soutoldold{technique}} have been extended to \esbaddold{\soutoldold{non-powder samples}single crystal systems}~\cite{schaub_exploring_2007}.
By collecting the full momentum transfer vector $\mathbf{Q}$ dependent intensity distribution, $I_{\textrm{tot}}(\mathbf{Q})$, from a single crystal and applying a FT, a quantity akin to $G(r)$ is obtained, \oldrobadd{namely the three-dimensional (3D)-PDF, $P_{\textrm{tot}}(\mathbf{r})$,}
which is proportional to the probability of finding two atom pairs separated by a \esbaddold{{\it vector}} distance $\mathbf{r}$.
This retention of directional information in $P_{\textrm{tot}}(\mathbf{r})$ can be advantageous if there is ambiguous overlap of features in the scalar function $G(r)$, but it is important to keep in mind that
\firstrevadd{$P_{\textrm{tot}}(\mathbf{r})$ and $G(r)$ are not perfect analogs.
Specifically, $P_{\textrm{tot}}(\mathbf{r})$ is most often defined in the literature as the FT of the diffracted intensity distribution $I_{\textrm{tot}}(\mathbf{Q})$~\cite{kobas_structural_2005,schaub_exploring_2007, weber_three-dimensional_2012}, whereas $G(r)$ is typically the FT of the reduced structure function $F(Q)$. 
Unfortunately this distinction can be overlooked, and although the 3D-PDF community could benefit from an extensive discussion of terminology along the lines of what has been done within the 1D-PDF community~\cite{Keen2001}, this lengthy task will not be tackled here.}

The more practical extension to the full 3D-PDF is the 3D-$\Delta$PDF technique~\cite{schaub_exploring_2007}, which relies on the ability to separate the $I_{\textrm{tot}}(\mathbf{Q})$ function into a sum of the Bragg component, $I_{\textrm{Bragg}}(\mathbf{Q})$, arising due to the long-range average structure \esbaddold{and leading to the well known Patterson function in direct space}, and the diffuse component, $I_{\textrm{diff}}(\mathbf{Q})$, arising due to local deviations from the long-range average structure, such that $I_{\textrm{tot}}(\mathbf{Q})=I_{\textrm{Bragg}}(\mathbf{Q})+I_{\textrm{diff}}(\mathbf{Q})$.
The 3D-$\Delta$PDF, $P_{\textrm{diff}}$ is then defined as
\begin{equation} \label{eq:fullPDF}
\begin{split}
P_{\textrm{diff}}(\mathbf{r}) & = FT[I_{\textrm{diff}}(\mathbf{Q})]\\
& = FT[I_{\textrm{tot}}(\mathbf{Q})-I_{\textrm{Bragg}}(\mathbf{Q})],
\end{split}
\end{equation}
or the FT of the full 3D intensity distribution $I_{\textrm{tot}}(\mathbf{Q})$ after subtracting out the Bragg component $I_{\textrm{Bragg}}(\mathbf{Q})$.
A full accounting of the theory behind these expressions can be found in earlier work~\cite{schaub_analysis_2011, weber_three-dimensional_2012}.
\firstrevadd{\soutoldoldold{It is important to note that}}The progression to 3D-$\Delta$PDF analysis was preceded in large part by earlier work studying the full 3D intensity distribution $I_{\textrm{tot}}(\mathbf{Q})$, without the application of a FT.
Much of the techniques of 3D-$\Delta$PDF analysis then arise as an extension \rjkadd{of} this previous work~\cite{epstein_least-squares_1983,weber_structural_2001, welberry_problems_2005,welberry_diffuse_2010} to the measurement \rjkadd{over broad\soutold{of extended}} $\mathbf{Q}$-ranges.

\firstrevadd{\soutoldoldold{
3D-$\Delta$PDF analysis has proven effective in a number of cases.
Early practical study demonstrating the technique have been largely qualitative in nature.
The first application of the technique identified the presence and approximate length scales of inter and intra- atomic cluster disorder and correlations in a Al–Co–Ni quasicrystal
A later qualitative work on the organic tricarboxamide revealed local ordering, where the authors found the probability of homochiral molecule pairs to be lower for direct neighbor molecules, and higher
for the second-order neighbors.
Subsequent work revealed superstructure columnar units in the quasicrystal Al$_{65}$Cu$_{20}$Co$_{15}$ having a diameter of about 14.5~\AA, with only weak lateral correlations.
This work also detailed some of the theoretical background behind 3D-$\Delta$PDF analysis, which were further explored in a later work exploring different explicit types of local effects in the 3D-$\Delta$PDF.
}}

\firstrevadd{3D-$\Delta$PDF analysis has proven effective in a number of cases, with
the earliest studies largely qualitative in nature.
The first application of the technique identified the presence and approximate length scales of inter- and intra-atomic cluster correlations in an Al–Co–Ni quasicrystal~\cite{kobas_structural_2005}.
Later qualitative works have revealed local ordering in tricarboxamides~\cite{schaub_exploring_2007}, and superstructure columnar units in the quasicrystal Al$_{65}$Cu$_{20}$Co$_{15}$~\cite{schaub_analysis_2011}.}

In a progression to a more quantitative analysis, short range ordering parameters were successfully refined from the 3D-$\Delta$PDFs~\cite{simonov_yell_2014,urban_real_2015}.
Later studies have revealed short-range ionic correlations in the intercalation compounds $\beta$'-Na$_{0.45}$V$_{2}$O$_{5}$~\cite{krogstad_reciprocal_2020} and local structure effects in the thermoelectric $\beta$-Cu$_{2-x}$Se~\cite{roth_solving_2019}.

While the usefulness of this relatively new technique may be clear, much of the methodology\esbaddold{, particularly related to data reduction,} remains murky and spread across various disparate works~\cite{kabsch_evaluation_1988,welberry_problems_2005,kobas_structural_2005,schaub_exploring_2007, weber_three-dimensional_2012,kabsch_processing_2014}.
Here we outline in extensive detail the process of 3D-$\Delta$PDF data collection and data processing, with the aim of making this new and powerful technique more accessible and transparent.
We adopt the \cis\ \esbaddold{cubic spinel material (space group \fd3m)} as a convenient test system, as its local structure has been previously studied using standard powder PDF, where a fluctuating orbital-degeneracy-lifted state was discovered, manifesting as a \esbaddold{subtle} local \esbaddold{symmetry breaking} distortion on the Ir pyrochlore sublattice~\cite{bozin_local_2019}\firstrevadd{\soutoldoldold{.
We utilize the existence of distortions as they warrant} and dictating} that the 3D-$\Delta$PDF signal will be non-zero.
We\rjkadd{\soutold{ should}} emphasize that the investigation of these distortions themselves is not the focus of this study, and will be addressed in a followup work.
\esbaddold{The \cis\ spinel\rjkadd{\soutold{ system}} is\rjkadd{\soutold{ further}} an exemplar ternary with appreciable scattering contrast between the constituent elements: Cu (Z=29), Ir (Z=77), and S (Z=16).
As will be seen from our data, pairwise correlations associated with the distortions can be observed in 3D-$\Delta$PDF even for pair vectors corresponding to the weakest scatterers in the system.}


Using this test system, we investigate the impact on the final 3D-$\Delta$PDF of data-processing parameters\oldrobadd{, including x-ray count time, detector artifact removal, background subtraction, \rjkadd{\soutold{absorption}interframe scale} correction, reciprocal space sampling, data merging, outlier removal, data symmetrization, Bragg intensity removal, and Fourier transformation.}
We \rjkadd{\soutold{find that}identify a number of situations where} the 3D-$\Delta$PDF is\rjkadd{\soutold{ itself}} extremely robust\rjkadd{\soutold{,}.} 
\rjkadd{In many cases, the 3D-$\Delta$PDF is} reproducible even with\rjkadd{\soutold{ what may seem like}} sub-critical data and data-processing.
\rjkadd{We also identify and discuss a few areas where caution must be exercised.}

\subsection{Organization}
\label{section:org}

\begin{figure}
\includegraphics[width=0.8\textwidth]{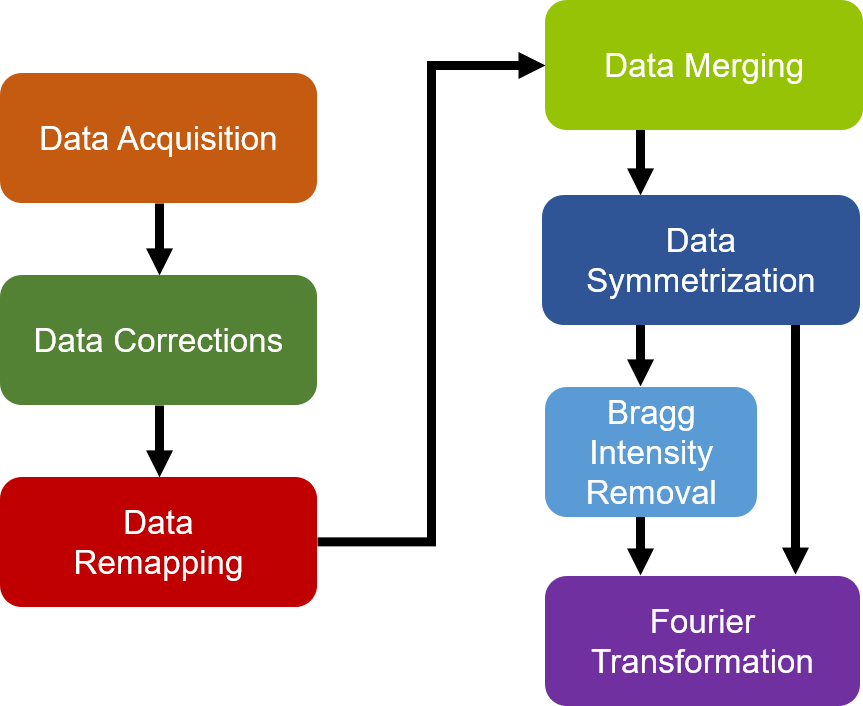}
\caption{General data work flow:
A flow chart depicting the main steps in producing 3D-PDF and 3D-$\Delta$PDF.
\rjkadd{\soutold{Measurement, }}Data corrections and remapping to reciprocal space must occur prior to subsequent steps.
Following this, data can be merged, symmetrized, and Fourier transformed (FT) to obtain 3D-PDF in any order.
For 3D-$\Delta$PDF, Bragg intensity removal must take place prior to the FT step.
\rjkadd{Outlier removal can be incorporated at the data remapping, merging, and symmetrization steps.}
}
\label{fig:flowchart}
\end{figure}

For practical purposes this paper is organized to follow the generic work flow progression involved in a 3D-PDF/3D-$\Delta$PDF experiment, detailed in Fig.~\ref{fig:flowchart}.
In section~\ref{section:meth} we outline details of the physical diffraction experiment and crystal samples, and give a brief overview of software \esbaddold{platforms} used in the process.
We then discuss in section~\ref{section:artifacts} pathological issues with the raw diffraction data brought about by detector artifacts \rjkadd{and sample imperfections}, as well as their correction.
In section~\ref{section:transform} we discuss the transformation of raw detector images from the detector frame of reference to crystal reciprocal space.
The steps described in subsequent sections~\ref{section:merge}-\ref{section:ft}, respectively on merging data from a single sample, applying symmetry operations to the observed signal, outlier removal, removing Bragg intensity, and applying a Fourier transform, can, with some restriction \esbaddold{that we address later},
be implemented in an arbitrary order.
The robustness of this process with regard to ordering and other factors is discussed in section~\ref{section:robustness}.
Wherever possible this presentation ordering is maintained, with  \rjkadd{a few\soutold{two}} exceptions.
\rjkadd{The procedure and effects surrounding outlier removal are discussed in section~\ref{section:outlier}, but outlier removal is included in all data reconstruction, merging, and symmetry averaging prior to this section.}
Occasionally the final Fourier transformed 3D-$\Delta$PDF is shown prior to section~\ref{section:ft} to discuss the effect of various reduction steps on the direct space data.
\oldrobadd{In addition, data remapped from detector space to reciprocal space are shown prior to section~\ref{section:trans} to highlight the reciprocal space extent of Bragg peaks.}

\section{Methods and Approaches}
\label{section:meth}
\subsection{Experimental Details}
\label{section:exp}
Single crystal diffraction measurements \firstrevadd{\soutoldoldold{at 300~K}} were carried out at the P21.1 beamline at the Positron-Elektron-Tandem-Ring-Anlage (PETRA III) facility at Deutsches Elektronen-Synchrotron (DESY), using an x-ray beam of 106 keV energy ($\lambda$ = 0.1170~\AA) sized to $0.5\times0.5$~mm$^2$.

\begin{figure}
\includegraphics[width=0.8\textwidth]{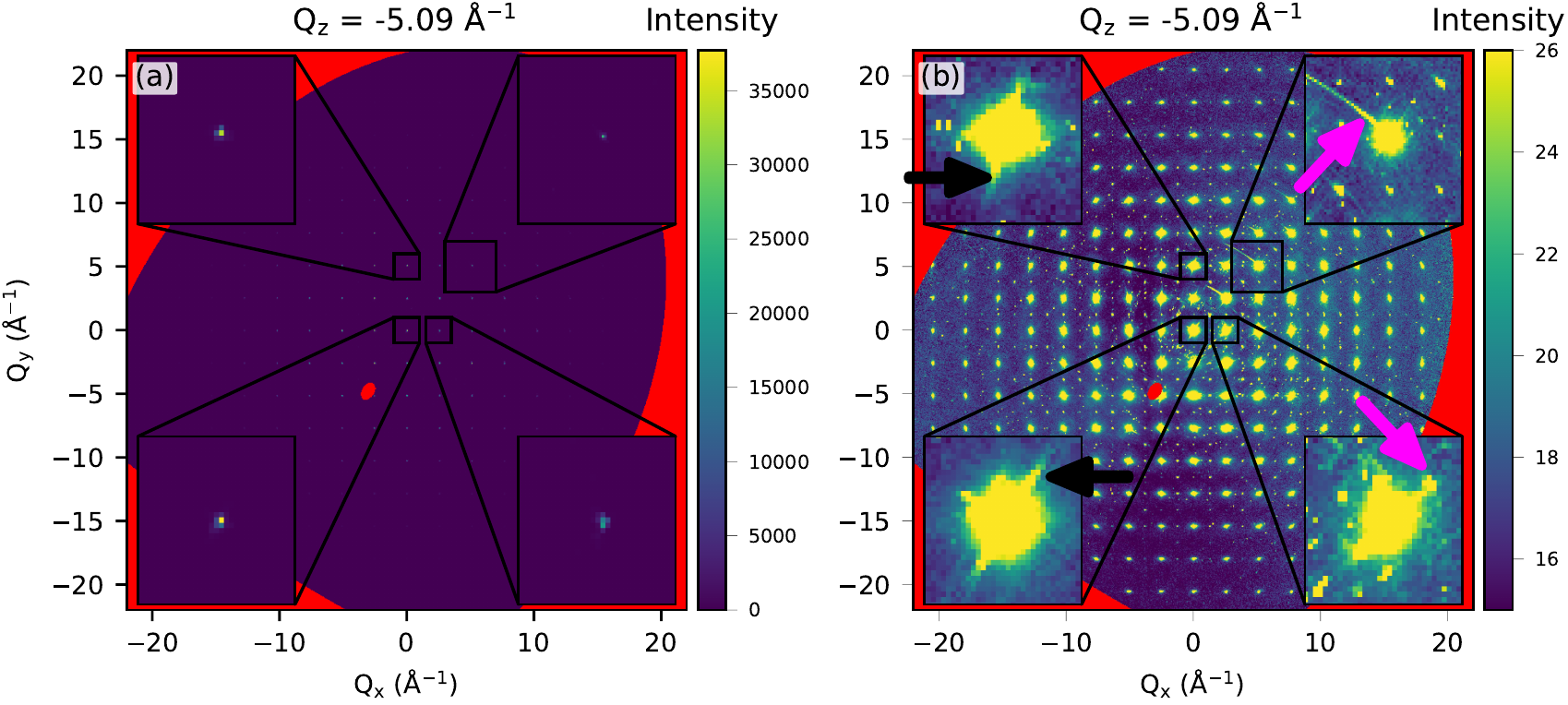}
\caption{Examples of Bragg reflections, blooming, and afterglow effects in data remapped from detector to crystal reciprocal space:
(a) An intensity map of a representative slice of reciprocal space with intensity scale chosen so as to highlight the Bragg reflections, shown in corner insets on an expanded scale.
\esbaddold{Notably, Bragg reflections span 1-3 reciprocal space voxels.}
(b) Identical to (a) with a distinct intensity scale chosen to highlight the diffuse features, as well as the pathological issues caused by unmitigated detector blooming and afterglow effects (marked by black and magenta arrows, respectively) described in the text.
\firstrevadd{\soutoldoldold{In situations where faithful observation of the diffuse signal is the goal,}}It is common for the Bragg features to be orders of magnitude stronger than both the detector artifacts and the diffuse signal, 
\firstrevadd{\soutoldoldold{This will of course}, although this can} vary from system to system.
}
\label{fig:bragg_peaks}
\end{figure}

Two octahedrally shaped single crystals \rjkadd{\soutold{($\sim$~2~mm lateral size)}} of \cis\ were measured.
\rjkadd{The first, herein referred to as `sample 1,' showed maximum dimensions of $\sim$~$720\times630\times520$~microns, while the second, herein referred to as `sample 2,' showed maximum dimensions of  $\sim$~$810\times750\times650$~microns, where the third listed dimensions are those parallel to the mounting and rotation axis.}
When utilizing a \rjkadd{flat panel area detector\soutold{image plate detector}}, the instrumental resolution function is heavily impacted by the projection of the beam or crystal footprint (whichever is larger) on the detector, and as such, achieving high spatial resolution requires that the beam or sample be as small as possible.
In this study since only relatively large crystals were available, the preference was to reduce the beam size.
This resulted in primarily isotropic Bragg peaks spanning about 0.13-0.19~\RAA\ in reciprocal space (Fig.~\ref{fig:bragg_peaks}(a)).
\firstrevadd{\soutoldoldold{Crystal morphology is also important, as highly anisotropic crystals will introduce a rotation-angle-dependent interframe scale fluctuation and/or instrumental resolution function.
This issue can be avoided if smaller crystals, on the order of 100 micrometers, are used, and care is taken that the crystal is centered within the beam.}}
\oldrobadd{Instrumental resolution width in $\mathbf{Q}$-space impacts the intensity decay of the (full) 3D-PDF as a function of $\mathbf{r}$ in direct space~\cite{weber_three-dimensional_2012}, specifically a \rjkadd{\soutold{larger}lower} instrumental resolution leads to a more rapid decay of the 3D-PDF \rjkadd{signal}.
It is an important aspect to consider, as one must be sure that the intensity decay of the (full) 3D-PDF extends beyond the spatial extent of any short-range distortions under investigation using the 3D-$\Delta$PDF.}

Each crystal was epoxy-mounted on the end of an amorphous cactus needle which was in turn mounted on a goniometer head \firstrevadd{\soutoldoldold{capable of 360\ignorespaces\textdegree\ rotation}}\rjkadd{, with the crystal carefully aligned to achieve adequate centering.}
\rjkadd{A well aligned sample implies that it resides at the center of rotation and the X-ray beam at all times.}
Measurements were carried out in air at ambient temperature and pressure.
Each diffraction image was collected with a PerkinElmer (PE) \rjkadd{1621} amorphous silicon flat panel detector ($2048\times2048$ pixels, $200\times200$~micron \rjkadd{pixel} size) located 516.6~mm \rjkadd{away} from the sample.
Detector distance, tilt, and rotation were calibrated using a CeO$_2$ standard measured in an identical geometry.
With this beam energy and geometry, the detector provides a 21~\ia\ range coverage of reciprocal space ($\qmax=21$~\ia).
\oldrobadd{Notably the \qmax\ value \rjkadd{\soutold{controls}determines} the resolution in direct space, $\Delta_{r}$ such that $\Delta_{r}=2\pi/\qmax$.
With the $\qmax=21$~\ia\ achieved here, $\Delta_{r}=0.3$~\AA.}

A single 3D measurement here consisted of an entire 360\ignorespaces\textdegree\ rotation of the crystal along one axis only, with the detector counting continuously and read out at regular sub-intervals of this rotation.
For each crystal, a total of 13 full 3D measurements were carried out, with exposure times per readout varying from 0.1 to 1.0~seconds and either 1800, 3600, or 7200 total images per full rotation (corresponding to 0.2, 0.1, or 0.05\ignorespaces\textdegree\ per image, respectively).
The benefit of considering such permutations of detector exposure time and angular image step size will be discussed in section~\ref{section:merge}.

\subsection{Software Platforms and Computation Details}
\label{section:software}
The vast majority of computational overhead in this work was handled with Python version 3.67~\cite{10.5555/1593511} and MATLAB version 9.6.0~\cite{MATLAB:2019}.
MATLAB was used to transform  raw detector images to reciprocal space, and to fill reciprocal space intensity after Bragg intensity removal.
Python was used for building dynamic detector masks, to find crystal orientation matrices, and to normalize, merge, symmetrize, \rjkadd{ \soutold{punch}remove Bragg peaks,} and Fourier transform reciprocal space intensity distributions.
Specifically, we made use of the NumPy, SciPy, Dask, and Matplotlib python packages~\cite{harris2020array,2020SciPy-NMeth,dask,Hunter:2007}.
The majority of the computations for this work were carried out on the Maxwell computational resources operated at DESY.
\subsection{Terminology}
\label{section:term}
\firstrevadd{\soutoldoldold{Prior to delving into the details of our analysis,}}It is useful to define some terminology used throughout the work which may be unfamiliar.
\oldrobadd{Diffracted intensity distributions \rjkadd{are considered} in either reconstructed crystal reciprocal space or detector space native to the measurement.}
In reconstructed crystal reciprocal space the coordinate chosen is momentum transfer vector $\mathbf{Q} = Q_x \mathbf{a}^* + Q_y \mathbf{b}^* + Q_z \mathbf{c}^*$ in units of~\RAA, where $Q_x$, $Q_y$, and $Q_z$ are continuous real numbers, and  $\mathbf{a}^*$, $\mathbf{b}^*$, and $\mathbf{c}^*$ are reciprocal lattice vectors.
We refer to discrete pieces of this space as `voxels' throughout this work.
In detector space the coordinate used is the position on the 2D detector.
Transformation between the two spaces is discussed in section~\ref{section:trans}.
We refer to discrete pieces of this space as `pixels' throughout this work.
\firstrevadd{\soutoldoldold{Considerations of differential ($\Delta$)}}Pair distribution functions are discussed in coordinates of $\mathbf{r} = x \mathbf{a} + y \mathbf{b} + z \mathbf{c}$ in units of~\AA, where $x$, $y$, and $z$ are continuous real numbers, and  $\mathbf{a}$, $\mathbf{b}$, and $\mathbf{c}$ are lattice vectors.
The process of removing \firstrevadd{and interpolating} Bragg intensity \firstrevadd{\soutoldoldold{and replacing it with diffuse intensity}} is colloquially called ``punch and fill," and this terminology has been adopted here.
Section~\ref{section:artifacts} deals primarily with the detector artifacts of blooming, where pixels adjacent to saturated pixels record erroneous intensity, and afterglow where saturated pixels record erroneous intensity in subsequent frames.
These two artifacts are a result of detector saturation, which for the purposes of this work is when the detector readout ceases to respond linearly with \oldrobadd{scattered} photon fluence.

\firstrevadd{When comparing data subject to differing processing procedures (section~\ref{section:robustness}), we have quantified the difference using $R_{\textrm{diff}}$, analogous to $R_{\textrm{split}}$  used in the field of serial crystallography~\cite{White2013},
\begin{equation} \label{eq:rdiff}
R_{\textrm{diff}} = \frac{1}{2^{1/2}}\frac{\sum \lvert I_1-I_2 \rvert}{ \frac{1}{2}\sum (I_1+I_2)}*100,
\end{equation}
where $I_1$ and $I_2$ are the individual, mutually valid data points of the two data-sets being compared.}

\section{Detector Artifacts}
\label{section:artifacts}

\begin{figure}
\includegraphics[width=0.8\textwidth]{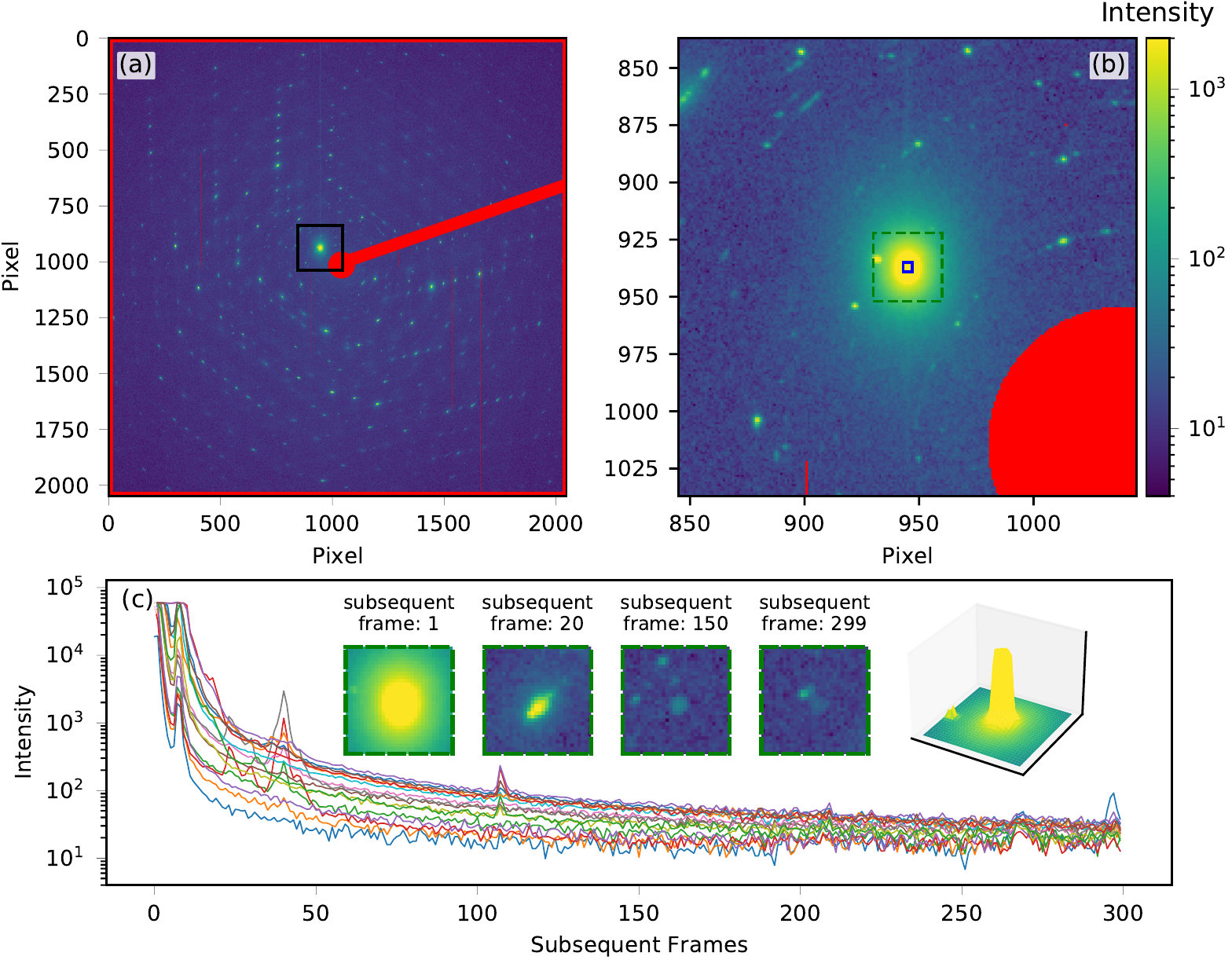}
\caption{Examples of blooming and afterglow effects in the raw experimental data:
(a) An intensity map of a detector image at the time point where a saturation event occurs, with a black box highlighting the location of the saturation/blooming.
(b) An expanded view\firstrevadd{\soutoldoldold{, marked by} of} the black square in (a).
(c) A plot of pixel intensity vs the number of images subsequent to the detector saturation event shown in (a) and (b).
Each plotted line in (c) represents one pixel from the \firstrevadd{\soutoldoldold{small }}blue square in (b).
Inset center in (c) are portions of diffraction images shown on an enlarged scale after the labeled number of subsequent frames, where the initial saturation event, featured in panels (a) and (b), occurs in frame zero.
Inset right in (c) is a 3D density plot of the portion of the detector subject to saturation, represented by the green dashed-line box in (b), at the time of saturation.
The truncation of this peak intensity indicates the detector has \firstrevadd{\soutoldoldold{exceeded its dynamic range,}} reached saturation, while the leakage of intensity into the surrounding detector area represents blooming.
In (a) and (b), red portions represent masking due to the beam stop and known bad pixels.
\esbaddold{Intensity peaks observable in panel (c) on top of decaying signal are subsequently detected diffraction events.}}
\label{fig:artefacts_raw_dat}
\end{figure}

The PE detector and other similar 2D detectors have many advantages for 3D-$\Delta$PDF measurements\firstrevadd{\soutoldoldold{, namely}.}
They are relatively inexpensive and as such are available at many hard energy X-ray beam lines, and they are relatively robust against permanent beam damage that could be caused by excessively strong Bragg intensities~\cite{perez-mendez_signal_1987}.
\firstrevadd{\soutoldoldold{Additionally,}} Such detectors often have high-sensitivity (detective quantum efficiency $ > 65\%$\ \rjkadd{at 80 keV}), relatively fast readouts (up to 30 Hz), and small point-spread functions \rjkadd{of less than one pixel}~\cite{chupas_applications_2007,lee_synchrotron_2008}.
These advantages have led to their widespread use within the powder total scattering or one-dimensional powder PDF field~\cite{chupas_rapid-acquisition_2003}.

\begin{figure}
\includegraphics[width=0.8\textwidth]{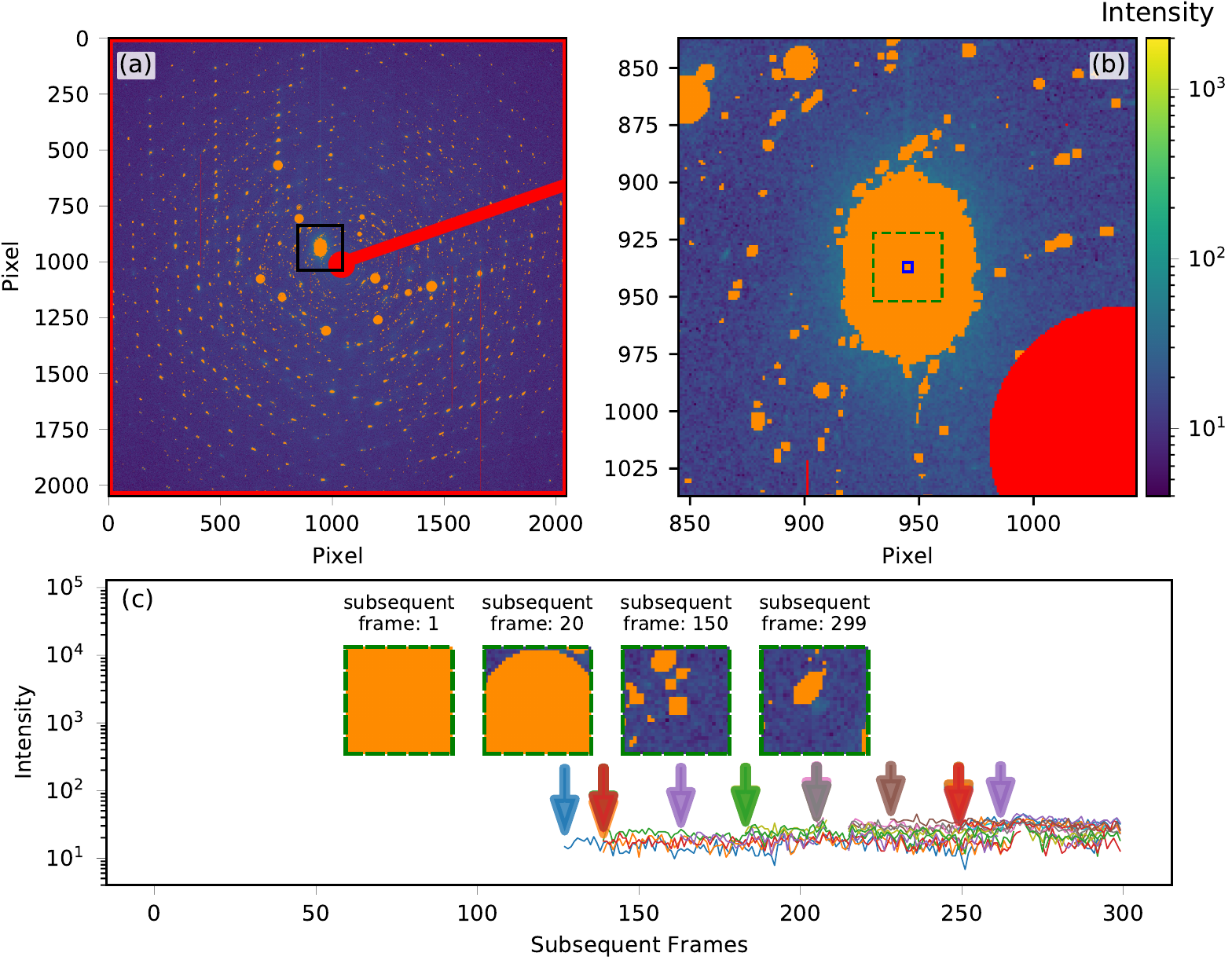}
\caption{Examples of the application of the dynamic masking heuristic adopted in this work:
(a) and (b) are identical to Fig.~\ref{fig:artefacts_raw_dat}(a) and (b), where raw data masked due to detector blooming and afterglow have been colored orange.
(c) Identical to Fig.~\ref{fig:artefacts_raw_dat}(c), where raw data masked due to detector blooming and afterglow have been omitted from the intensity plot.
Colored arrows in (c) indicate the frame at which a given pixel becomes unmasked, and colors correspond to the line colors plotted in (c).
\esbaddold{Note that, in the detector region affected by the saturation event, the pixels do not fully recover until at least 150 frames after the saturation event}, with all pixels recovering by 275 frames after the saturation event.
}
\label{fig:dyn_mask_raw_dat}
\end{figure}

However, such detectors also suffer from a number of drawbacks.
The nature of the 3D-$\Delta$PDF measurement entails collecting the full diffraction signal from a single crystal specimen, with the intention of faithfully reproducing the diffuse intensity.
This necessitates that very intense Bragg spots will be incident on the detector.
With a dynamic range on the order of $10^3$, these Bragg spots will frequently saturate such detectors.
Under saturation, such detectors will exhibit two phenomena, \firstrevadd{\soutoldoldold{commonly}herein} referred to as afterglow \firstrevadd{\soutoldoldold{(or detector lag) }}and blooming\firstrevadd{\soutoldoldold{ (or cascading pixel spillover)}}.

Afterglow can be considered a forward temporal cross talk between detector frames collected at different times, whereby a saturated pixel or group of pixels in one read-out persists to read out an erroneously elevated count rate for a finite number of detector frames \firstrevadd{\soutoldoldold{including and }}following the saturation event.
It is a result of the finite discharging time required of the storage capacitance in the sensor layer of the detector~\cite{albagli_performance_2005,chupas_applications_2007}.
Effectively, portions of the sensor layer retain charge in subsequent image read-outs.
Detector blooming~\cite{welberry_problems_2005} typically occurs when saturated pixels overflow excess charge into neighboring pixels, causing a readout of erroneous intensity from these spatially adjacent pixels.

Examples of these two detector artifacts \firstrevadd{\soutoldoldold{as they manifest }}in \firstrevadd{\soutoldoldold{raw}} detector images are shown in Fig.~\ref{fig:artefacts_raw_dat}.
While the given geometry should produce Bragg peaks spanning approximately 5-10 detector pixels, it is clear from Fig.~\ref{fig:artefacts_raw_dat}(b) that the Bragg peak causing saturation of the detector leads to elevated counts in the surrounding 60 detector pixels, evidenced by the halo of intensity surrounding the Bragg peak.
Further, plotting the intensity of the saturated pixels as a function of the subsequent frame number, shown in  Fig.~\ref{fig:artefacts_raw_dat}(c), reveals that the saturated pixels continue to readout elevated counts for at least 150 subsequent frames.

Importantly, both artifacts introduce undesirable erroneous intensity into the measurement, causing issues in the intensity distribution when this is transformed into crystal reciprocal space, as shown in Fig.~\ref{fig:bragg_peaks}(b).
Blooming creates a roughly equiaxed area of bright pixels surrounding the saturated pixel(s) within a detector image.
Since each detector image intersects the reciprocal space origin, a disk of intensity in detector space corresponds to a disk in reciprocal space sitting on a surface which intersects the reciprocal space origin. 
In 2D slices of reciprocal space, detector blooming manifests as streaks of intensity sitting on arcs which are oriented towards the reciprocal space origin, as indicated by black arrows in the upper and lower left insets of Fig.~\ref{fig:bragg_peaks}(b).

Detector afterglow is a persistence of elevated counts in the saturated pixel(s) across multiple subsequent frames.
The scalar \textit{magnitude} of the momentum transfer vector of each pixel in the detector is fixed by the detector position and orientation, and does not change as the crystal is rotated in lab-coordinates.
The \textit{direction} of the momentum transfer vector associated with each detector pixel is however altered by the crystal rotation.
As a result, detector pixels affected by afterglow are mapped in reciprocal space to bright arcs around the reciprocal space origin, with constant momentum transfer magnitude.
In 2D slices of reciprocal space, detector afterglow manifests as arc-like streaks of intensity centered at the reciprocal space origin, such as those seen in the upper right inset of \ref{fig:bragg_peaks}(b) (features indicated by magenta arrows).
\firstrevadd{The reciprocal space orientations of both blooming and afterglow artifacts are determined by the crystal orientation and rotation axis.}

Disentangling the effects of detector blooming and afterglow from the true diffuse intensity distribution can be difficult, but are essential to a successful experiment, as they can lead to sub-critical reciprocal space data.
Previous strategies for handling detector blooming and afterglow in single crystal scattering measurements have simply omitted effected portions of the detector~\cite{welberry_problems_2005}.
In this work, we adopted a heuristic whereby we identified detector saturation along with the subsequent afterglow and blooming within the raw detector images through a simple intensity threshold approach of 40,000 counts.
The pixels surpassing the intensity threshold were masked within the subsequent 200 frames to reduce afterglow.
To reduce blooming, pixels within a diameter of 40 pixels encompassing any saturated pixel were masked in the initial frame, and this diameter was decreased linearly in the \rjkadd{subsequent 100 frames\soutold{subsequent frames 100}}.
By processing an entire image sequence using this heuristic prior to remapping to reciprocal space, an evolving or dynamic mask of excluded pixels was built and subsequently applied to each measured detector image during reciprocal space remapping.
\firstrevadd{\soutoldoldold{
We found the above-listed masking parameters represent a point of diminishing returns in terms of correcting for detector artifacts.
That is, adopting a lower threshold and/or a larger diameter results in significant masking of regions not impacted by detector blooming/afterglow, without providing a justifiable improvement  in actual blooming/afterglow effects.
}}

\begin{figure}
\includegraphics[width=0.8\textwidth]{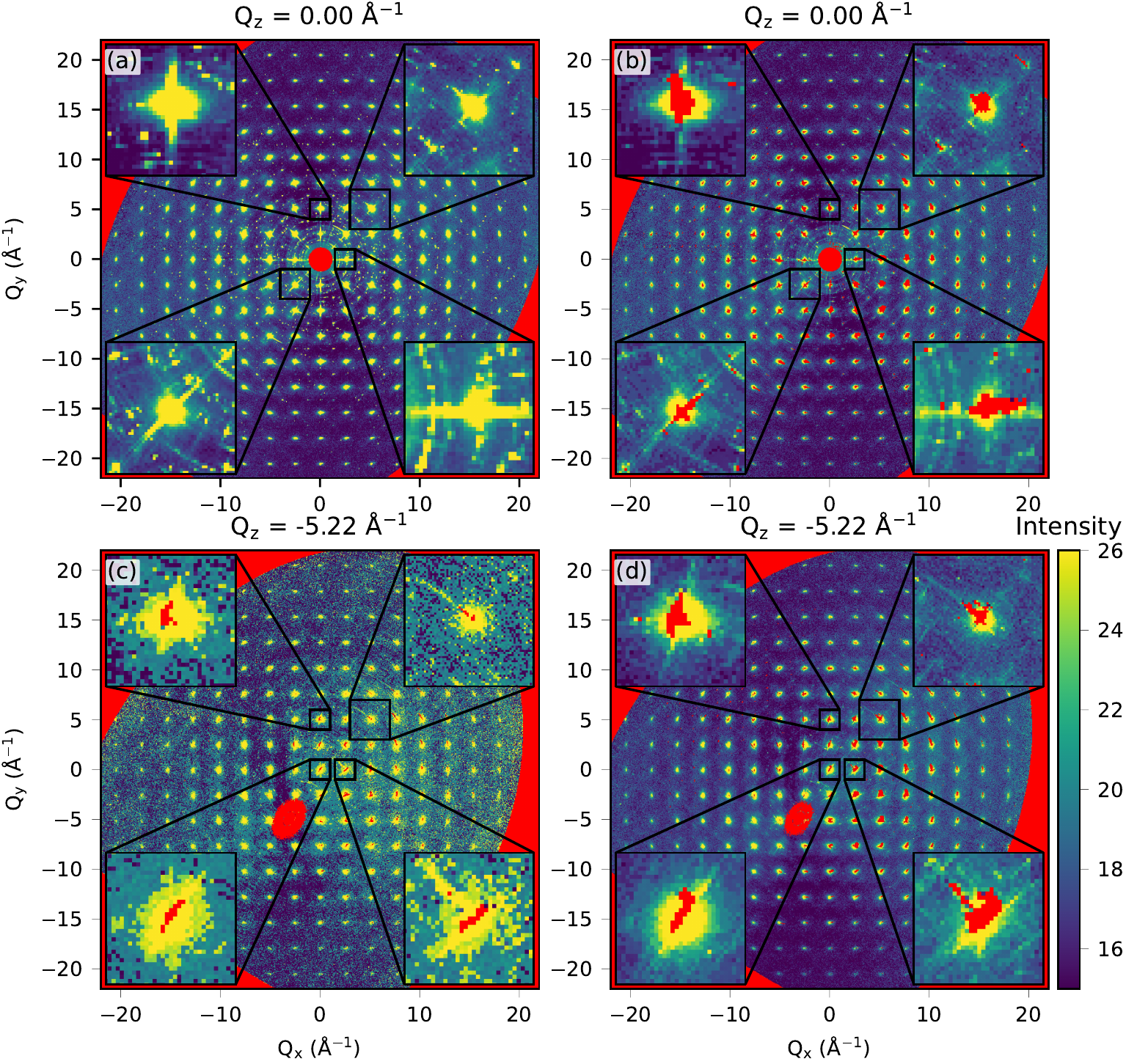}
\caption{Examples of the impact of dynamic masking in the data remapped to reciprocal space:
Intensity maps of representative slices of reciprocal space \rjkadd{(a) without} and \rjkadd{(b, c, d) with} the dynamic masking heuristic described in the text to mitigate the effects of detector blooming and afterglow.
\rjkadd{$Q_z$ slices were chosen to best demonstrate the effect, and are maintained throughout many of the figures.}
Images used in the reconstruction for (c) were counted for 0.1 seconds per frame, and those in \rjkadd{(a, b, d)} were counted for 0.8 seconds per frame.
A few regions heavily impacted by the masking, with missing data \esbaddold{eliminated by masking marked} in red, are shown inset on an enlarged scale.
Note that the counting time, contrasted \firstrevadd{\soutoldoldold{between}in} (c) and (d) heavily impacts the \firstrevadd{\soutoldoldold{degree to which detector saturation manifests within the data and subsequent}extent of} dynamic masking\firstrevadd{\soutoldoldold{, and also impacts}} and overall noise.
\esbaddold{Shorter counting time results in less masking but noisier data, while longer counting improves the statistics but requires more masking.}
}
\label{fig:dyn_mask_recon_dat}
\end{figure}


In Fig.~\ref{fig:dyn_mask_raw_dat} we demonstrate the application of such dynamic masking heuristic to the detector images.
Dynamically masked pixels are shown in panels (a) and (b) in orange, and are excluded from the intensity plot in panel (c).
The dynamic masking excludes all saturated pixels, along with a large portion of the surrounding detector area in the initial frame.
In subsequent frames (shown inset in panel (c)) the original masked area shrinks until it is removed entirely, unless a subsequent saturation event occurs nearby.
Fig.~\ref{fig:dyn_mask_raw_dat}(c) demonstrates that the dynamic masking effectively excludes afterglow intensity until the detector has recovered to baseline count rates, with recovery beginning and ending approximately 150 and 275 frames after the saturation event, respectively.

The dynamic masking heuristic we outline here uses the raw detector images as an input to generate image-wise masks, which are then used in reconstruction to reciprocal space.
The processes leads to the exclusion of \firstrevadd{\soutoldoldold{erroneous}} portions of the measured data, \firstrevadd{\soutoldoldold{which can yield}producing empty} regions of reciprocal space \firstrevadd{\soutoldoldold{which contain no data}}\rjkadd{ if these voxels are not successfully measured in other detector images.}
Examples of dynamically masked reciprocal space reconstructions are presented in Fig.~\ref{fig:dyn_mask_recon_dat},
\firstrevadd{\soutoldoldold{It is worth noting that these plots represent}using} a significantly abbreviated color scale to highlight the \firstrevadd{\soutoldoldold{subtle contrast between the}weaker} features \firstrevadd{of interest.\soutoldoldold{in question and Bragg peaks, which themselves are 3 orders of magnitude more intense (see e.g. Fig.~\ref{fig:bragg_peaks})}.}
\firstrevadd{\soutoldoldold{With this in mind, it can be seen by}Reconstructions with and without masking are shown in} Fig.~\ref{fig:dyn_mask_recon_dat}\rjkadd{(a) and (b)}\firstrevadd{, respectively, and demonstrate that} that dynamic masking removes a considerable amount of the strongest effects due to detector saturation, blooming, and afterglow.
\rjkadd{The lower insets in these panels, focusing on areas near the reciprocal space origin, also reveal that \firstrevadd{\soutoldoldold{dynamic }}masking reduces the prominence of what can be considered parasitic scattering contributions.
Composed of a collection of bright spots, these ring-like features are not expected in a single crystal measurement, and likely are a result of small polycrystalline inclusions and/or side-crystals.
Originating from sample imperfections, this situation yields what could be considered sub-critical scattering data, but can be remedied to a great extent through the dynamic masking process.}

\rjkadd{Although dynamic masking does aid in removing detector artifacts,} some streaking does persist in the reciprocal space intensity maps.
\oldrobadd{While it may seem that this could be remedied by a more aggressive dynamic detector masking, we found that the reported dynamic detector masking parameters and the associated results represents an optimal point, beyond which we see diminishing returns.
That is, adopting a more aggressive approach results in significant masking of regions not impacted by detector blooming/afterglow, without providing a justifiable improvement in actual blooming/afterglow effects.}
\rjkadd{As will be seen in subsequent sections, further processing of the reconstructed data, including merging (section~\ref{section:merge}) and symmetrization (section~\ref{section:symm}) significantly mitigate the effects of detector artifacts which survive dynamic masking, especially when including outlier rejection (section~\ref{section:outlier}).}

The quantity and size of the masked regions in reciprocal space are dependent on the parameters of the masking heuristic and the degree of afterglow and blooming, which is related to the exposure time per frame.
As can be seen in Fig.~\ref{fig:dyn_mask_recon_dat}(c) and (d) respectively, short exposure times results in less masking and fewer/smaller \rjkadd{completely} masked regions, while longer exposure times results in more masking and more/larger \rjkadd{completely} masked regions.
We found that across all data-sets, each comprised of a full rotation of the crystal, dynamic masking with the parameters we utilized resulted in a mean pixel rejection rate of about 3\%, with a maximum rejection rate per image of about 7\%, a relatively small fraction of the information content of each image.

As will be seen in our subsequent discussions, the missing portions of reciprocal space data (holes in the data) created by this limited dynamic masking \rjkadd{are nearly\soutold{can be}} completely filled (recovered) \rjkadd{after merging\soutold{through}} data-sets collected over distinct exposure time and angular mesh (section~\ref{section:merge}) and by applying crystal symmetry (section~\ref{section:symm}).
Thus, any compromised/excluded signal of relevance in reciprocal space can be recovered by the oversampling that is achieved in detector space over the course of the measurement \firstrevadd{and data processing}.
Further, many of these holes exist at positions of Bragg peaks, which are punched and filled later in the data analysis pipeline (section~\ref{section:punchfill}).
\rjkadd{We can conclude that with a relatively high-symmetry crystal and extensive data over-sampling}, such a dynamic masking heuristic and the resulting apparent gaps in individual data-sets should not represent any significant issue for the analysis of the final treated data.
\firstrevadd{In situations where the crystal shows lower symmetry, it may be necessary to recover masked intensity (achieve over-sampling) by utilizing more than a single rotation axis during measurement, as the orientation of detector artifacts is dictated by the rotation axis.}

\section{Transforming From Detector to Reciprocal Space}
\label{section:transform}

\begin{figure}
\includegraphics[width=0.8\textwidth]{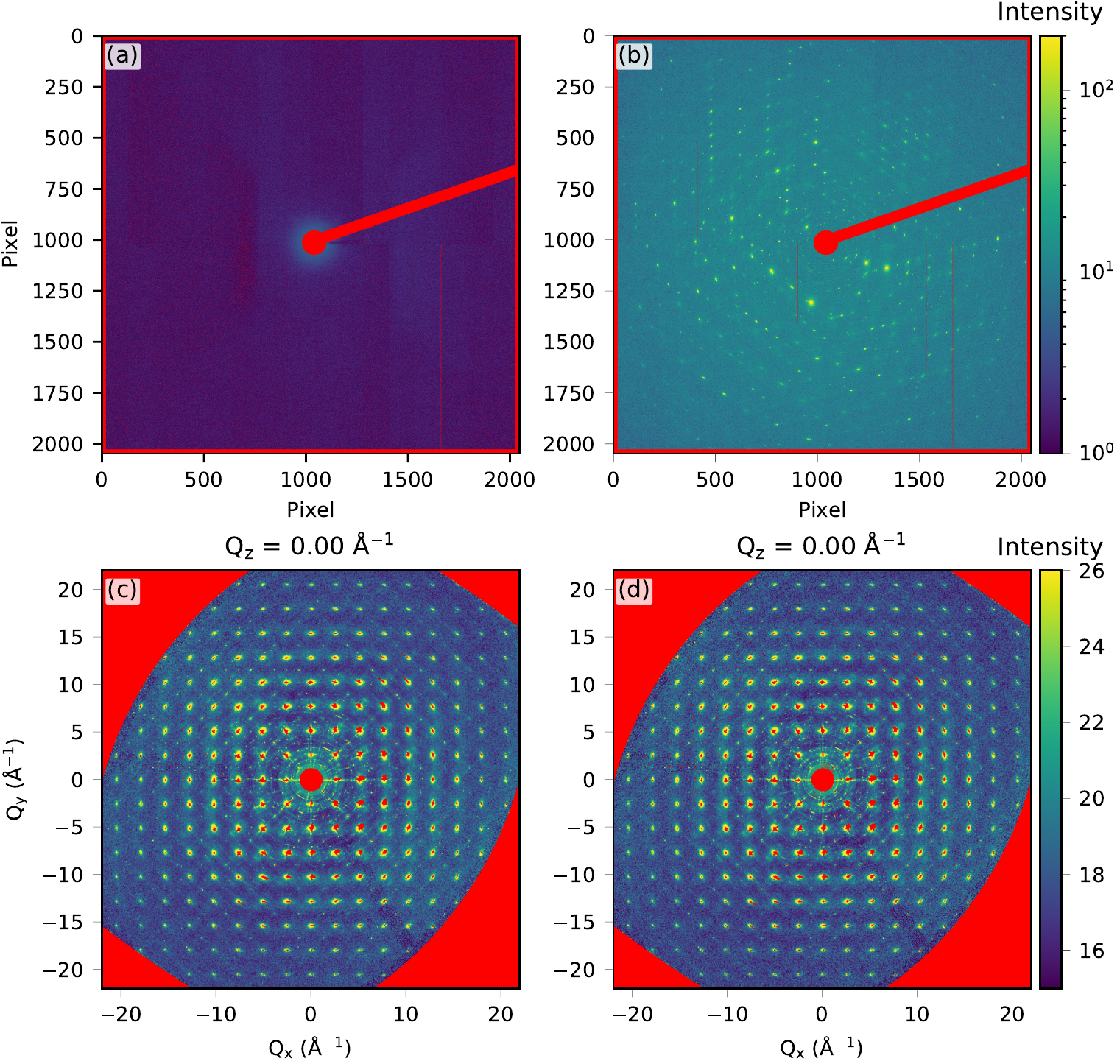}
\caption{The impact of experimental background on  data remapped to reciprocal space:
(a) The background diffraction signal measured for this experiment.
The data are relatively featureless aside from a small region surrounding the direct beam.
(b) A raw detector image after background subtraction.
(c) An intensity map of a representative slice of reciprocal space with background subtraction.
(d) The same slice of reciprocal space without background subtraction.
In our example, the relatively strong signal from the crystal dominates, with very little contribution from the background.}
\label{fig:bkg}
\end{figure}
\subsection{Background Subtraction}
\label{section:bkg}
\firstrevadd{As with 1D-PDF experiments, it is important in a 3D-$\Delta$PDF experiment to ensure that only coherent scattering from the sample under study contributes significantly to final reduced 3D-$\Delta$PDF.
Contributions not due to coherent scattering from the sample can include air scattering, scattering from the sample mount or housing, or incoherent contributions from the crystal, such as fluorescence and Compton scattering.
First priority should be given to minimizing the contribution of each  extraneous scattering source.
Failing this, these contributions must be removed or their effect taken into account during analysis of the final 3D-$\Delta$PDF.}

If the diffraction measurement is carried out in air, or if the crystal is mounted in such a way that additional material is present in the beam (such as epoxy or a mounting stub) and the background component does not vary as a function of crystal rotation, it is likely simplest to conduct a pixel-wise subtraction of a single, \rjkadd{averaged}, background image collected from an identical setup (excluding the crystal sample).
\rjkadd{If the crystal absorbs a large portion of the primary beam, it is possible that the air scattering occurring between the crystal position and the beam stop shows a substantial difference between when the crystal is present or absent.
This can be remedied by minimizing air scattering and using a crystal with low absorption, or failing this, a scale parameter can be fit to minimize the background signal~\cite{holm_temperature_2020}.}
If the background has many components (both air and epoxy, for example), it may be necessary to independently subtract a scaled diffraction pattern of each isolated component.
It is best practice to minimize this component of the background signal as much as possible, e.g. by carrying out the experiment in low scattering chamber, \rjkadd{minimizing the beam-path in air (collimator and beam-stop close to crystal)}, or ensuring that no components of the mounting fixture are in the X-ray beam.

\firstrevadd{Incoherent contributions to the background, such as fluorescence and Compton scattering from the crystal, cannot be remedied by subtracting the measurement of an empty sample environment.
The relative fraction of these two contributions compared to the coherent signal should be optimized by utilizing suitably chosen high-energy x-rays~\cite{Ramsteiner2009} and/or energy discriminating detectors, if available~\cite{broennimann_pilatus_2006,henrich_experience_2008}.}

\firstrevadd{There are a number of approaches within 3D-$\Delta$PDF studies to handling background.
Examples are fitting the number background, including the incoherent contributions, with a smooth functions for subtraction~\cite{holm_temperature_2020}, assuming certain portions of reciprocal space contain no diffuse signal and interpolating between these regions~\cite{schaub_exploring_2007, weber_three-dimensional_2012, simonov_experimental_2014}, or taking the floor (minimum) counts of averaged detector frames to build a reciprocal space map of background for subsequent subtraction~\cite{krogstad_reciprocal_2020}.
A number of works avoid mentioning incoherent contributions explicitly, presumably implicitly grouping fluorescence and Compton scattering into a general background~\cite{schaub_exploring_2007, schaub_analysis_2011,krogstad_reciprocal_2020,roth_solving_2019}.
As fluorescence and Compton scattering from the sample both are smooth and vary slowly in reciprocal space as a function of scalar $Q$ rather than vector $\mathbf{Q}$, their contributions should bias scale factors during quantitative modeling/fitting of the 3D-$\Delta$PDF.
In such situations, imposing additional physically reasonable constraints for pair correlations can help to resolve this bias~\cite{weber_three-dimensional_2012, simonov_experimental_2014}.
Neglecting fluorescence and Compton scattering entirely can provide adequate data for identifying gross features and trends in the 3D-$\Delta$PDF~\cite{roth_solving_2019}, and has also yielded meaningful quantitative results ~\cite{krogstad_reciprocal_2020}}

In the current work, a diffraction pattern of an unloaded sample environment was used as background, and subtraction was done pixel-wise for each detector image.
In our particular experiment, the background signal from an unloaded sample environment was several orders of magnitude smaller than that of the sample, and primarily concentrated at the reciprocal space origin (Fig.~\ref{fig:bkg}(a)).
In this case, background subtraction does not result in a significant difference in the raw detector image (Fig.~\ref{fig:bkg}(b)) or the reciprocal space intensity distributions (Fig.~\ref{fig:bkg}(c) and (d)).
This may not hold true \rjkadd{in situations where the magnitude of the background scattering is comparable to that of the sample  e.g.} for less crystalline samples, \rjkadd{physically smaller crystals}, or if an in-situ diffraction cell contributes significant background.


\subsection{\rjkadd{\soutold{Absorption Correction}}Interframe Scale Correction}
\label{section:absorp}
\begin{figure}
\includegraphics[width=0.8\textwidth]{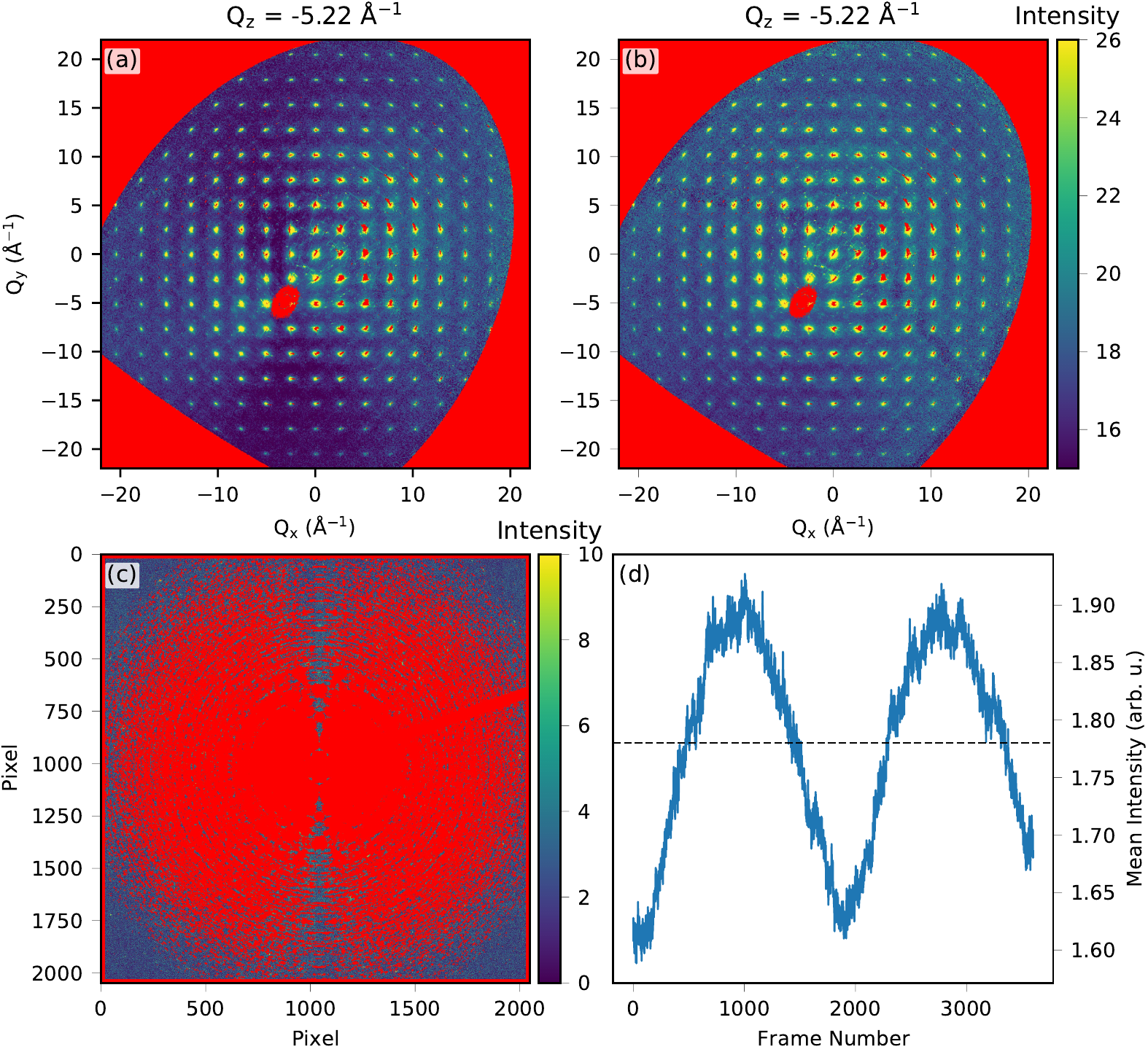}
\caption{The impact of sample \rjkadd{\soutold{absorption}interframe scale fluctuation} on reciprocal space data:
(a) An intensity map of a representative slice of reciprocal space prior to applying an \rjkadd{\soutold{absorption}interframe scale correction}.
Note the dark cones in the intensity distribution.
(b) The same data as in (a), after applying the \rjkadd{\soutold{absorption}interframe scale} correction described in the text.
\oldrobadd{(c) An example portion of the detector used in correcting for \rjkadd{\soutold{absorption}interframe scale fluctuations}, with excluded pixels masked in red.
(d) The mean intensity considering all non-masked pixels in (c) for each frame in a data-set, as well as the mean across the entire data-set, represented by a horizontal dashed line.
The dark cones seen in (a) have been replaced by a more uniform intensity distribution in (b) as a result of scaling each image such that frame-to-frame variation seen in (d) is eliminated.}
}
\label{fig:absorb}
\end{figure}

\firstrevadd{\soutoldoldold{If the crystal is relatively large, significant absorption of the X-ray beam may occur, and if the crystal morphology is not perfectly equiaxed, the amount of crystal in the beam, and thus the absorption 
and the overall amount of scattering, may vary as a function of the crystal rotation angle.
Absorption specifically can be exacerbated if the combination of sample chemistry, beam energy, and crystal size together result in significant absorption, although this situation should generally be avoided by selecting adequate beam energy and utilizing a suitably small crystal with an isotropic morphology, if such an ideal crystal is available.
If the crystal dimensional cross section shows a smooth variation with rotation angle, a smooth variation in diffracted intensity will be observed across the series of images measured while rotating the crystal.
Within the data remapped to crystal reciprocal space, this is manifested as dark cones observable in the background intensity distribution, which can be seen in  Fig.~\ref{fig:dyn_mask_recon_dat} and Fig.~\ref{fig:absorb}(a).}}

\firstrevadd{If the crystal is relatively large and the morphology is not perfectly equiaxed, a variation in diffracted intensity will be observed across the series of images measured while rotating the crystal. This occurs as the volume of the crystal in the beam changes during rotation, impacting the absorption and amount of scattering.
Within the data remapped to crystal reciprocal space, this is manifested as dark cones observable in the background intensity distribution, which can be seen in  Fig.~\ref{fig:dyn_mask_recon_dat} and Fig.~\ref{fig:absorb}(a).
This extent of this variation can generally be minimized by selecting adequate beam energy and utilizing a suitably small crystal with an isotropic morphology, if such an ideal crystal is available.}

To mitigate the impact of these frame-to-frame fluctuations \rjkadd{in the current study} we have adopted an empirical approach used in previous work~\cite{welberry_problems_2005}.
We identify the portion of the detector which does not show any diffraction peaks when considering a full crystal rotation data-set,
\oldrobadd{ utilizing the dynamic masks generated in section~\ref{section:artifacts}.
That is, all pixels of the detector never subject to a dynamic mask are used.
A representative example of this portion of the detector is shown in Fig.~\ref{fig:absorb}(c).}
The mean pixel intensity across this portion of the detector is then computed individually for each detector image in a full data-set.

\oldrobadd{In Fig.~\ref{fig:absorb}(d) we show an example of the frame-wise mean intensity for the detector region shown in Fig.~\ref{fig:absorb}(c), along with the overall mean for the entire data-set (all frames), represented by a horizontal dashed line.
To remove the interframe intensity fluctuations, each image is scaled prior to reconstruction to reciprocal space by the ratio of the frame mean to the overall mean for a given data-set such that the variation in this mean intensity from frame to frame is eliminated.}

An example of a reciprocal space intensity distribution before and after such correction can be seen in Fig.~\ref{fig:absorb}(a) and Fig.~\ref{fig:absorb}(b), respectively.
The application of this crystal orientation dependent, frame-wise \rjkadd{\soutold{absorption}interframe scale correction} successfully removes the dark cones in the intensity distribution, leading to a more uniform intensity distribution.

\subsection{Geometric Data Remapping Protocol}
\label{section:trans}
Following data collection and correction for pathological detector related issues, each data-set is transformed from detector to crystal reciprocal space.
This is carried out using information such as the calibrated beam energy, detector position and orientation, \rjkadd{ unit cell dimensions\soutold{crystal structure} \esbaddold{of the sample}}, and known crystal rotation step size between detector images \esbaddold{used in the experiment for a given data-set}.

Effectively, the reciprocal space momentum transfer vector for each pixel in an image is computed, and if this pixel is not subject to masking, the measured intensity, corrected for polarization effects~\cite{milch_indexing_1974,kabsch_evaluation_1988,zachariasen_theory_1994, kabsch_processing_2014}, is binned into the appropriate voxel in reciprocal space, and the reciprocal space voxel bin count, representing the total number of pixels contributing to a given voxel, is incremented.
Once each image for a given measurement \esbaddold{(data-set)} is processed, the voxel bin is \oldrobadd{normalized (divided) by its bin count} and written to a file.
The voxel bin counts can also be written out, as they can subsequently be used as bin weights in further transformations.

\subsection{Reciprocal Space Sampling}
\label{section:samp}
\begin{figure}
\includegraphics[width=0.8\textwidth]{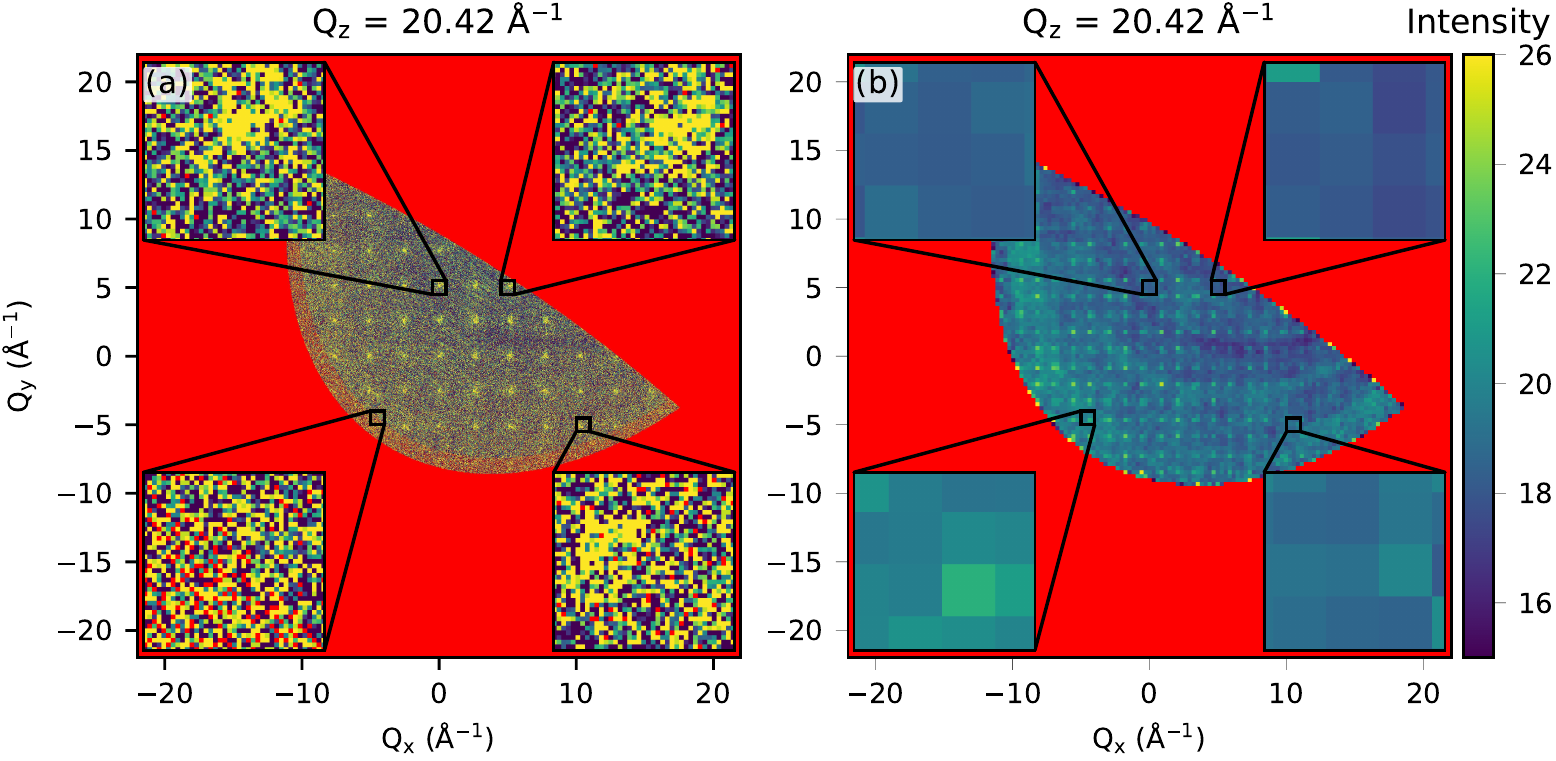}
\caption{Choice of reciprocal space sampling grid:
(a) An intensity map of a representative slice of reciprocal space \firstrevadd{\soutoldoldold{where the remapping from detector space has been done }}with a step size of 0.026~\RAA\ ($1701\times1701\times1701$ voxels).
Note that, within the inset regions on an enlarged scale, portions of reciprocal space inside the maximum momentum transfer vector of the experiment contain no data (represented by red pixels among colored pixels.)
(b) Similar to (a), with a step size to 0.29~\RAA\ ($151\times151\times151$ voxels).
The insets demonstrate that diffuse features and Bragg peaks now span the same 1-3 voxels, and would be difficult to disentangle during the punch and fill process.
}
\label{fig:sampling}
\end{figure}

During the remapping from detector to reciprocal space, \esbaddold{which transforms the units from physical pixels used in the detector space to reciprocal space units,} one must choose a voxel grid in reciprocal space on which to map the detector space pixels.
The choice of this grid is important, as it can impact both the quality of the result, as well as the computational overhead associated with the data processing.
A grid which is too coarse will impart a graininess to the reciprocal space intensity distribution, impede the punching and filling of Bragg peaks, and \rjkadd{\soutold{lead to a degraded resolution in the}will limit the field of view of the} 3D-$\Delta$PDF, \rjkadd{possibly producing issues with aliasing}.
An example of under-sampling in reciprocal space is shown in Fig.~\ref{fig:sampling}(b).
Here, a reciprocal space grid step size of 0.29~\RAA\ was chosen, equating to a $151\times151\times151$ voxel grid.
It is clear that, with such a coarse voxel grid, Bragg peaks spanning 0.19~\RAA are spread across a larger portion of reciprocal space, overlapping significantly with diffuse peaks.
This could potentially lead to issues during removal of Bragg features in subsequent steps.

A grid which is too fine leads to separate but equally problematic situations.
Fine voxel grids can produce arrays which are too large to handle easily.
If we assume a single voxel bin requires 16 bytes or memory, a grid such as the one used throughout the majority of this work ($701\times701\times701$ voxels, step size of 0.063~\RAA) requires about 5.5 gigabytes (GB) of memory and/or disk space.
Decreasing the step size to 0.026~\RAA\ ($1701\times1701\times1701$ voxels) increases the array size by a factor of more than 10, to almost 80 GB, which can become problematic if multiple arrays are to be handled simultaneously (as in merging and applying symmetry operations).
A too fine grid can also entail regions of reciprocal space inside the measured $Q$-range that have not been sampled by the detector, leaving holes in the transformed reciprocal space intensity distribution.
This can be seen in Fig.~\ref{fig:sampling}(a), where a reciprocal space grid of $1701\times1701\times1701$ corresponding to a grid step size to 0.026~\RAA\ voxels was used.
The figure insets show unsampled points, represented by red pixels, dispersed within regions of the intensity map which are within the measured $Q$-range.
\rjkadd{To avoid such under-sampling, the voxel size should approximately match the portion of reciprocal space spanned by pixels on the outer edges of the detector.}

\subsection{Crystal Orientation}
\label{section:orient}
Our experiments were carried out on a single crystal of unknown orientation, with known \rjkadd{\soutold{crystal structure}unit cell dimensions}.
Knowledge of the crystal orientation is not required to move from the detector reference frame to \esbaddold{\soutoldold{a general} an arbitrary} reciprocal space reference frame, as the latter is determined only by the detector position and orientation and the beam energy.
It is however useful and common to transform to a \textit{specific} reciprocal space reference frame whereby the reciprocal lattice of the crystal has a known (often orthogonal) relationship to the principal Cartesian axes.
To achieve this, \rjkadd{knowledge of the unit cell dimensions and space group is required\soutold{a \esbaddold{\soutoldold{complete}full} understanding of the average structure of the crystal under investigation is required}}, and the crystal orientation matrix must be determined (in this case after the measurement is completed).

There are numerous strategies for determining a crystal orientation matrix~\cite{kabsch_solution_1976,kabsch_evaluation_1988,kabsch_processing_2014}.
\esbaddold{\soutoldold{In our work}Here}, we applied the Kabsch algorithm~\cite{kabsch_solution_1976}, choosing to find the crystal orientation using \rjkadd{\soutold{the}difference vectors with coordinates corresponding to the} \{440\} family of Bragg peaks.
\firstrevadd{\soutoldoldold{In principle, any family of Bragg peaks can be used, provided they are strong and can be resolved from other peaks given the resolution of the experiment.
The \{440\} family was chosen here as it has a large multiplicity (12) in the  \fd3m\ space group describing this system at ambient conditions, it is well resolved, and corresponds to a long momentum transfer vector.
The last criterion provides that any absolute errors in estimating the momentum transfer vector during the crystal orientation correspond to smaller relative errors.}}

\firstrevadd{}
This process was carried out \esbaddold{independently for each set} of the full crystal rotation measurements done in this study, where we varied exposure time and crystal rotation step size.
The crystal orientation matrices were then used for each transformation from detector space to reciprocal space, such that the reciprocal lattice vectors of the crystal are parallel to the principal Cartesian axes.
\firstrevadd{\soutoldoldold{It is worth noting that }}The condition that reciprocal lattice vectors are parallel to the principal Cartesian axes is only possible with orthogonal crystal systems, as is the case here.
\firstrevadd{\soutoldoldold{Additionally}}One could transform into the crystal coordinates of non-Cartesian systems. 
This would of then dictate that intensity space array indices no longer correspond to a Cartesian axis, and additional care is required when e.g. plotting. 
This scenario can however be advantageous for certain data processing steps, such as symmetrization

\section{Merging Data}
\label{section:merge}

\begin{figure}
\includegraphics[width=0.8\textwidth]{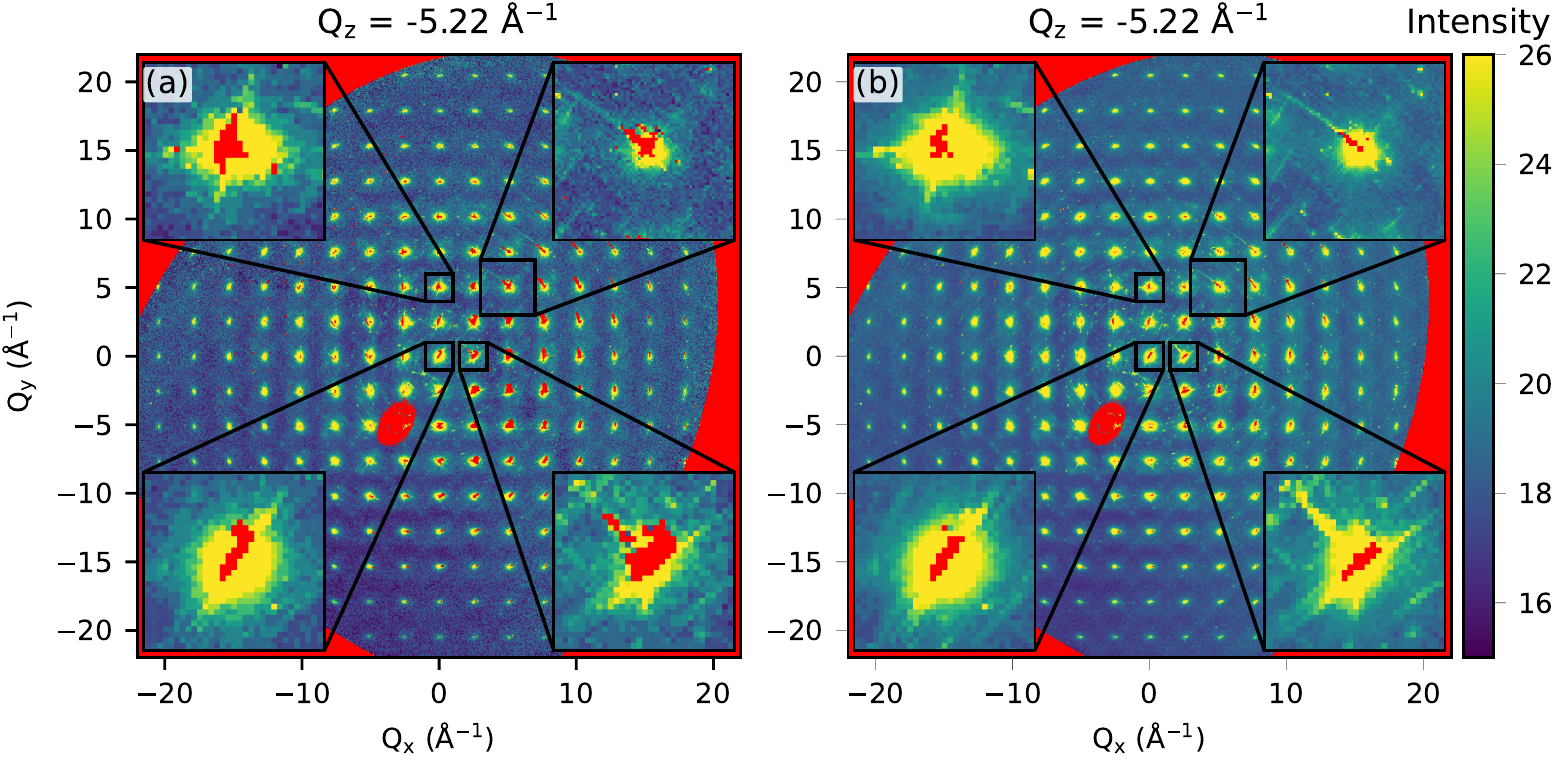}
\caption{Merging data collected at different exposures:
Intensity maps of a representative slice of reciprocal space of (a) a single data-set (1 s exposure per frame) and (b) data after merging 13 full data-sets collected from the same sample with differing exposure times.
A few regions heavily impacted by the merging process are shown inset on an enlarged scale, with missing data in red.
Merging multiple data-sets reduces anomalies due to detector blooming and afterglow, impacting the streaks highlighted in  Fig.~\ref{fig:bragg_peaks}(b).
Also, note that the merging process has partially filled in the missing intensity introduced by the dynamic masking process (see e.g. Fig.~\ref{fig:dyn_mask_recon_dat})
}
\label{fig:merged}
\end{figure}

Complete data-sets collected from the same crystal can be merged into a single data-set at any point in the data pipeline \textit{after} the transformation from detector to reciprocal space has been completed.
In practice this requires computing a weighted arithmetic mean of the intensity of each voxel bin across all relevant full data-sets, where the weights are \esbaddold{given by} the total number of detector pixels contributing to each voxel as discussed earlier.

Merging data-sets collected with different exposure times and angular grid meshes defined by the crystal rotation step size improves statistics, reducing the scatter of the overall data-set, which can arise from electronic glitches \rjkadd{or} shot noise\rjkadd{\soutold{, or even small impurity crystals present on the main crystal}}.
Merging also \esbaddold{facilitates filling in of the} portions of reciprocal space \esbaddold{that were} excluded due to detector saturation and the associated blooming and/or afterglow.
Shorter exposure times (see e.g. Fig.~\ref{fig:dyn_mask_recon_dat}(c)) produce less afterglow and blooming, and thus more faithfully reproduce the intensity distribution closer to Bragg peaks, at the expense of undercounting weak features far from Bragg peaks.
Conversely, longer exposure times (see e.g. Fig.~\ref{fig:dyn_mask_recon_dat}(d)) produce more afterglow and blooming, and lead to more and larger holes in the reciprocal space intensity distribution, but also more faithfully reproduce weak features far from Bragg peaks.

In Fig.~\ref{fig:merged} we show an example reciprocal space slice (a) before and (b) after merging 13 full data-sets collected from a single crystal with various exposure times and crystal rotation steps.
The holes introduced by our dynamic detector masking heuristic (see e.g. Fig.~\ref{fig:dyn_mask_recon_dat}) have been filled in to a large extent, and the regions between Bragg peaks show less scatter after merging.
\oldrobadd{Note that some very weak \rjkadd{\soutold{powder-ring like}} features are observable within reciprocal space slices \rjkadd{which are inconsistent with the lattice of \cis.}
\rjkadd{These are} likely due to small quantities of impurity polycrystalline inclusions \rjkadd{and/or side crystals of the primary \cis\ phase}.
\rjkadd{Many of these reside at fixed $Q$ values and thus constitute powder ring-like features.
The presence of such parasitic scattering contributions is best avoided by a careful selection of the crystal, as they can propagate to the 3D-$\Delta$PDF and cause spurious features.}
As will be seen these are remedied to a large degree by the data processing \oldrobadd{in later steps.}\rjkadd{\soutold{, and do not impact the features \esbaddold{\soutoldold{expected}} in the final 3D-$\Delta$PDF, which are expected to be isolated in direct space, rather than ring-like.}}}

\section{Symmetrization}
\label{section:symm}

\begin{figure}
\includegraphics[width=0.8\textwidth]{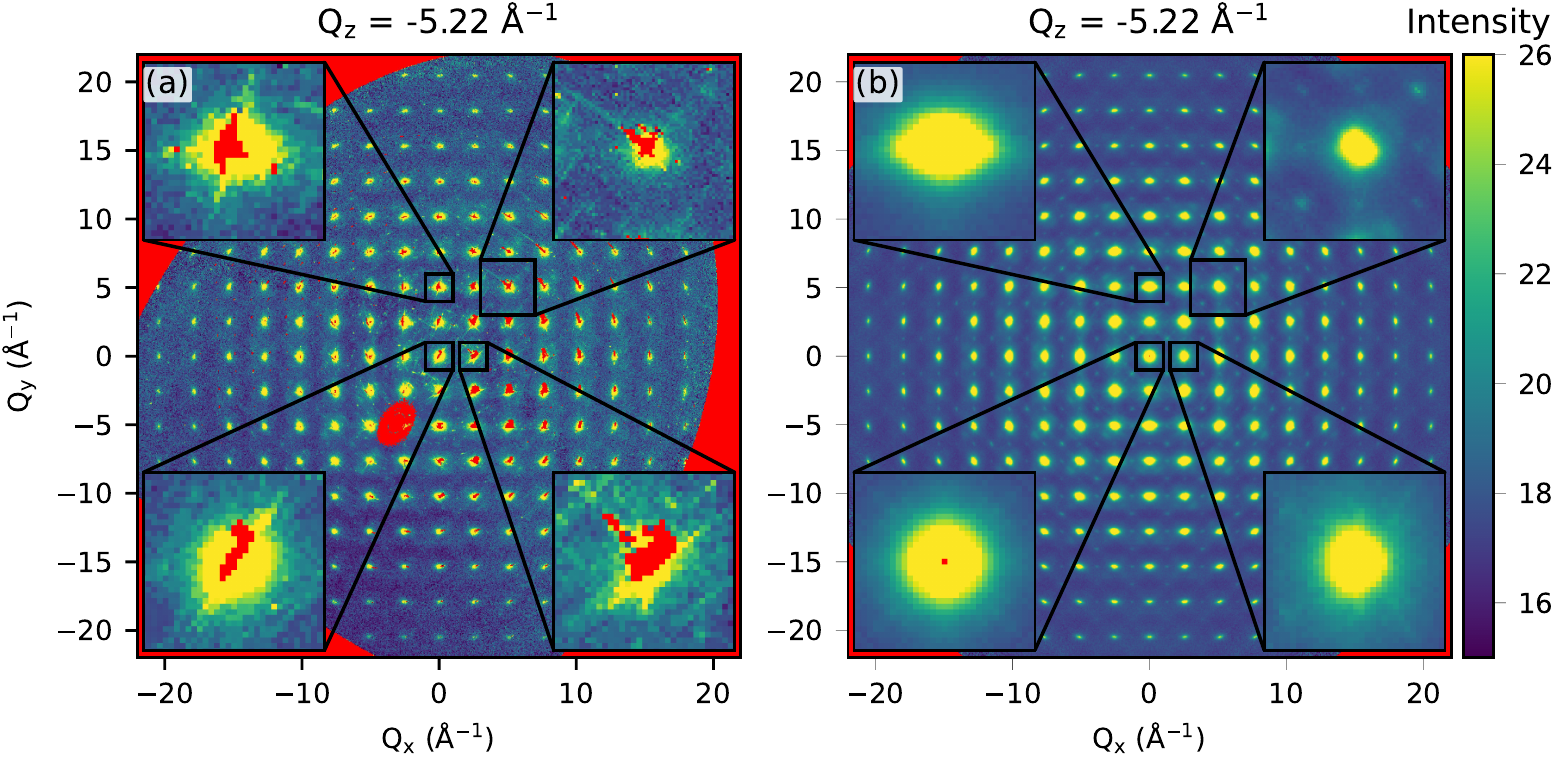}
\caption{Applying symmetry operations:
Intensity maps of a representative slice of reciprocal space of a single data-set (1 s exposure per frame) (a) before and (b) after averaging over all relevant symmetry operations.
A few regions heavily impacted by the symmetry averaging process are shown inset on an enlarged scale, with missing data in red.
The process of applying symmetry operations has filled in the majority of the missing intensity introduced by the dynamic masking process (see e.g. Figs.~\ref{fig:dyn_mask_recon_dat} and~\ref{fig:merged}).
\oldrobadd{A small hole does remain in the lower right inset of panel (b), but this occurs at a Bragg position and is filled in subsequent steps.}
The symmetry averaging has also removed all residual streaking associated with detector afterglow/blooming}
\label{fig:symm}
\end{figure}

With some exceptions,
the full intensity distribution (Bragg + diffuse) measured from the crystal should obey the same Laue point group symmetry governing the long-range average structure of the crystal~\cite{weber_structural_2001, welberry_diffuse_2010}.
\oldrobadd{This may not be the case if, for example, the crystal shows structural heterogeneity over length scales of the same order of magnitude as the beam footprint~\cite{weber_structural_2001}.
\rjkadd{\soutold{In the \cis\ case investigated here}For the \cis\ crystal used here}, this was not the case, as the beam footprint was relatively large, and the results are robust between individual single crystals.}
We can apply the symmetry operators associated with this Laue point group to the data under consideration.
This \esbaddold{symmetrization \soutoldold{serves to}} improves statistics, reduces the scatter of the data-set, fills in portions of reciprocal space which were excluded due to detector saturation and the associated blooming and/or afterglow, and also fills in portions of the space which were not measured due to limited crystal rotation.

In the \cis\ case, the Laue symmetry is $m\overline{3}m$, which contains 48 symmetry operations.
The application of each symmetry operation requires computing a weighted arithmetic mean of the intensity of each voxel bin across all relevant symmetry operators, where the weights are the total number of detector pixels contributing to each voxel.
\oldrobadd{The set of all symmetry equivalent voxels and their weights are obtained by applying the Laue symmetry to the $(HKL)$ coordinate of each voxel and voxel weight.}
Fig.~\ref{fig:symm} shows a representative slice of a reciprocal space intensity distribution (a) before and (b) after applying all relevant symmetry operators to a data-set which had been previously merged as described in the previous section.
The process of applying symmetry operators\rjkadd{ with outlier removal} has nearly completely filled in the vast majority of holes associated with the our dynamic masking heuristic, and has resulted in a \esbaddold{\soutoldold{much}substantially} cleaner and more uniform intensity distribution.
\oldrobadd{The small quantity of remaining holes (see e.g. Fig.~\ref{fig:symm}(b)) lower left inset) are a result of dynamic masking, and located at Bragg positions filled during the punch and fill process.
In addition, portions of reciprocal space which were not measured due to crystal orientation and detector geometry (outside a given $\mathbf{Q}$-range) have been filled in \rjkadd{by \soutold{my}} the symmetrization process.}
\firstrevadd{In situations where fewer symmetry operators are applicable, it may be necessary to achieve the required level of over-sampling through repeated measurement of the crystal utilizing distinct rotation axis.}

\section{Outlier Removal Protocol}
\label{section:outlier}

\begin{figure}
\includegraphics[width=0.8\textwidth]{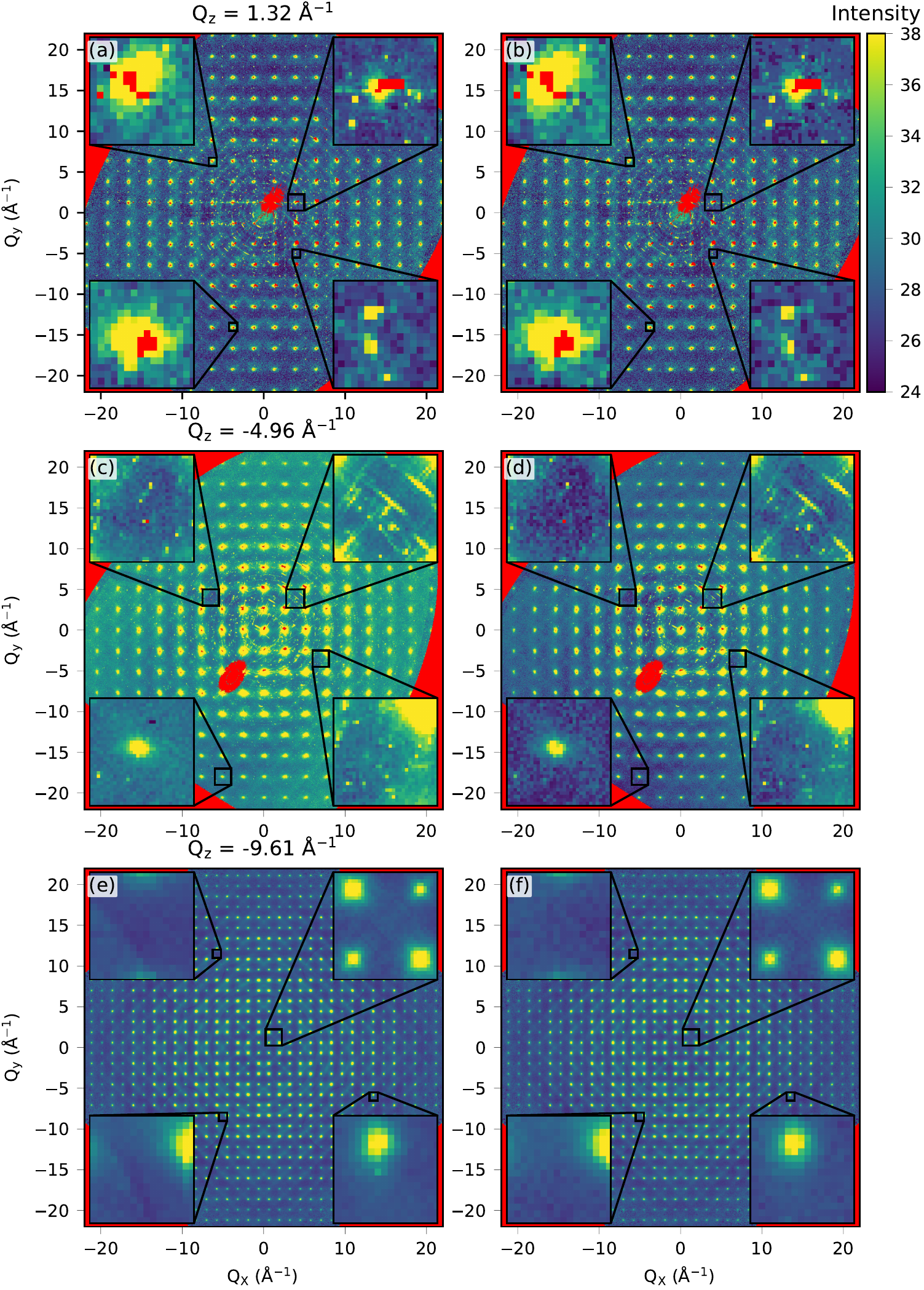}
\caption{Outlier Removal:
Intensity maps of a representative slice of reciprocal space at different stages of data processing without (a, c, e) and with (b, d, f) outlier removal for (a, b) the process of reconstructing from detector to reciprocal space, (c, d) merging data with different exposure times and crystal rotation step sizes, and (e, f) applying symmetry averaging.
Each plot contains insets showing enlarged scales for features of interest.}
\label{fig:outliers}
\end{figure}

In situations where data are \esbaddold{prone to \rjkadd{systematic errors\soutold{noise}}\soutoldold{noisy}} it is often useful to implement some method of outlier removal, where individual observations that are distant from the mean are excluded.
\esbaddold{\soutoldold{In this work}In 3D-$\Delta$PDF experiments}, \rjkadd{systematic errors\soutold{data noise}} can arise due to detector blooming/afterglow, small crystal impurities, and/or shot noise in the detector.
\rjkadd{Outlier removal has been successfully applied in 3D-$\Delta$PDF data reduction protocols previously, specifically at the symmetry averaging step~\cite{sangiorgio_correlated_2018, holm_temperature_2020}}
\esbaddold{\soutoldold{In the current work}Here}, we have \rjkadd{utilized \soutold{tested}} outlier removal at three distinct steps of data processing: when transforming from detector to reciprocal space, during merging of distinct data-sets, and during symmetry averaging.
Outlier removal effectively entails first computing the mean intensities and their standard deviation  for each voxel bin in reciprocal space, and then subsequently recomputing the means after the outlier values are removed using an outlier exclusion criteria.
During each of the above-mentioned steps of data processing we have chosen to exclude\esbaddold{, as an outlier,} any observation which falls outside a two standard deviations window around the mean.
\rjkadd{We have elected to use the mean rather than the median~\cite{blessing_outlier_1997} such that the \firstrevadd{\soutoldoldold{computational overhead is tractable.}required system memory and computation times are reduced by an order of magnitude to tractable ranges.
Whenever possible the median should be preferred, as the mean can be strongly biased if there are strong outliers.}}
The impact of outlier removal is distinct at each of the three averaging steps where it was applied.

Example slices of reciprocal space, where the data transformation from detector to reciprocal space was done with either \esbaddold{just} standard averaging or with outlier removal are shown in Fig.~\ref{fig:outliers}(a) and (b) respectively.
\oldrobadd{The panel insets show some moderate improvements.
\rjkadd{The $Q_z$ of each slice was chosen to demonstrate the maximum difference between outlier removal/inclusion.}
The primary effect is to decrease the intensity of some features which may be considered spurious.
At the data reconstruction step, data-points subject to outlier removal are individual detector pixel intensities.
Thus outlier removal serves to screen out pixels which are inconsistent with the mean of associated voxel to which the pixel is mapped.
For this reason, outlier removal at the data reconstruction step does not create large regions of differences when comparing (a) and (b), but rather isolated individual voxel changes.
}

Shown in Fig.~\ref{fig:outliers}(c) and (d) are the effects of merging data \firstrevadd{\soutoldoldold{collected with distinct exposure times or crystal rotation step sizes}} with either standard averaging or with outlier removal averaging, respectively.
In this case, the insets show a significant difference, with the intensity of many streak-like features diminished after outlier removal.
These features are associated with detector blooming/afterglow, and their diminished prevalence in Fig.~\ref{fig:outliers}(d) suggests that outlier removal during data merging offers a considerable benefit.

Lastly, Fig.~\ref{fig:outliers}(e) and (f) show the effect of symmetrization with either standard averaging or with outlier removal, respectively.
\oldrobadd{
The panel insets show moderate improvement; those on the left show a remedy of dark regions arising from data-set edge effects (regions around the beam-stop which create holes in reciprocal space).
The upper right panel shows a subtle reshaping of diffuse peaks, while
the lower right panel shows the removal of a spurious peak-like feature in the vicinity of a diffuse peak.
}

\section{Removing Bragg intensity}
\label{section:punchfill}

\begin{figure}
\includegraphics[width=0.8\textwidth]{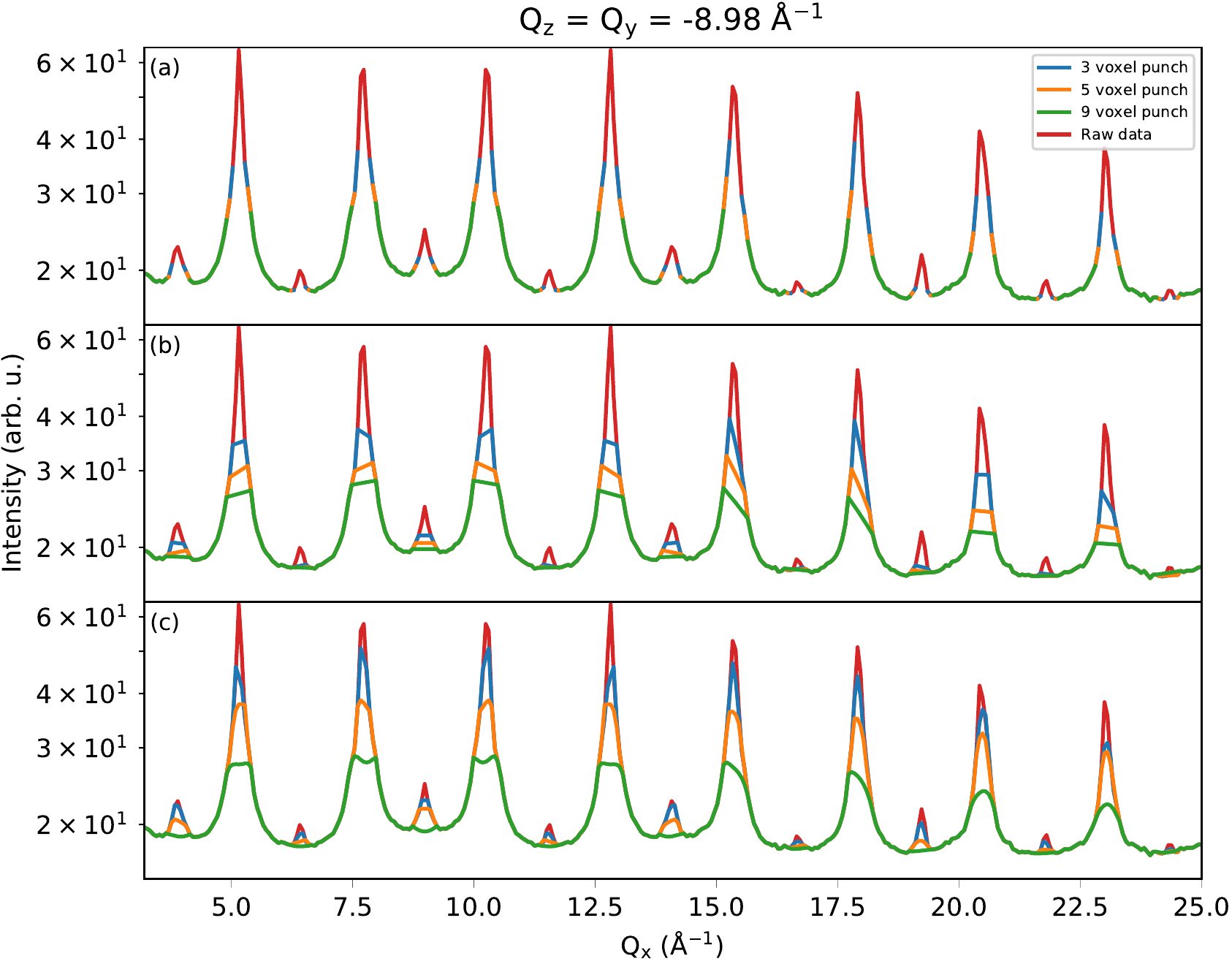}
\caption{Bragg punch size \esbaddold{as seen from a line scan perspective}:
(a) A line scan through a slice of reciprocal space intensity, demonstrating the effect of punch sizes or 3, 5, or 9 voxels spanning 0.189~\RAA, 0.315~\RAA, or 0.567~\RAA, respectively.
The associated linear interpolation (b) or DCT (c) filling.
Note that all data are plotted on a log scale to highlight both Bragg and diffuse features.
The plotted raw data represents reciprocal space intensity, without punching.
}
\label{fig:punch_line}
\end{figure}

\firstrevadd{\soutoldoldold{At this point in the reduction process (see Fig.~\ref{fig:flowchart}) there are two possible directions leading to direct space data.
The first possibility is to compute total 3D-PDF by performing a Fourier transform of the reciprocal space data described in section~\ref{section:ft}.
The second possibility is to separate the components of scattering corresponding to the ordered average structure (Bragg scattering) and to the deviations from it (diffuse scattering) leading to 3D-$\Delta$PDF.}}

\firstrevadd{\soutoldoldold{It is noteworthy that commonly used experimental 1D-PDF based on powder total scattering always represents a total PDF, and that observation of local structure deviations from the average requires utilization of adequate average structure model.
In contrast, simply obtaining a 3D-$\Delta$PDF requires a partial structural model which describes at minimum the long range average unit cell dimensions and symmetry.
To then obtain quantitative information on the true local structure, one must fully understand the long range average structure to which the 3D-$\Delta$PDF is referenced.
In this situation, information on the local structure can be, at least in principle, directly obtained from the 3D-$\Delta$PDF data in a manner independent of any model description of the local structure, provided the atomic pair vectors associated with the long range average structure can be identified in 3D space.
A further discussion of preliminary analysis is presented in section ~\ref{section:cisspecific}.}}

\firstrevadd{At this point in the reduction process (see Fig.~\ref{fig:flowchart}) there are two possible directions leading to direct space data~\cite{weber_three-dimensional_2012}.
The first is to compute total the 3D-PDF by performing a Fourier transform of the full reciprocal space data,and the second is to separate the components of scattering corresponding to the ordered average structure (Bragg scattering) and to the deviations from it (diffuse scattering), leading to 3D-$\Delta$PDF.}

Computing the 3D-$\Delta$PDF from the reciprocal space intensity distribution requires removal of any Bragg peak contribution while retaining all diffuse \esbaddold{\soutoldold{peak} signal contributions}.
This is typically called ``punch and fill" and a number of procedures have been outlined for the process, \firstrevadd{including filling with average intensity values~\cite{kobas_structural_2005}, convolution based filling adopted from the astrophysics community~\cite{krogstad_reciprocal_2020}, and even a structure-model independent approach based on statistical outlier detection~\cite{weng_k-space_2020}}.

Practically speaking, ``punch and fill" involves locating Bragg peaks, removing them from the data, and filling in any and all diffuse intensity removed by punching, which can include both \esbaddold{\soutoldold{diffuse} broad} and sharp features.
\esbaddold{The filling process effectively represents an attempt to best compensate for the removed part of the diffuse signal at the locations of punched Bragg peaks.}
It can be carried out in the raw detector images, or within the reciprocal space intensity distribution, but \firstrevadd{\soutoldoldold{all}most} techniques require knowledge of the crystal orientation\rjkadd{, unit cell dimensions, and space group,\soutold{and crystal unit cell size and shape}} as this information is necessary and sufficient to locate and remove all \esbaddold{(and only)} Bragg peaks.
Within this work, we have elected to operate on the reciprocal space intensity distribution although this choice does not change the generality.
\esbaddold{\soutoldold{In practice}Generally}, any three-dimensional shape can be used to remove Bragg intensity from the data.
In situations where the instrument resolution leads to Bragg peaks which are significantly anisotropic in crystal reciprocal space (or detector space)~\cite{weber_three-dimensional_2012}, it may be useful to adopt an anisotropic punch shape.
In this study, we found that Bragg peaks, which were on average about three orders of magnitude more intense than the observed diffuse features, were largely isotropic, spanning 1-3 voxels in reciprocal space (see e.g. Fig.~\ref{fig:bragg_peaks}).
\esbaddold{\soutoldold{As such,}For this reason here} we adopted an isotropic \rjkadd{spherical} voxel punch\esbaddold{\soutoldold{in the present study}}.

\begin{figure}
\includegraphics[width=0.8\textwidth]{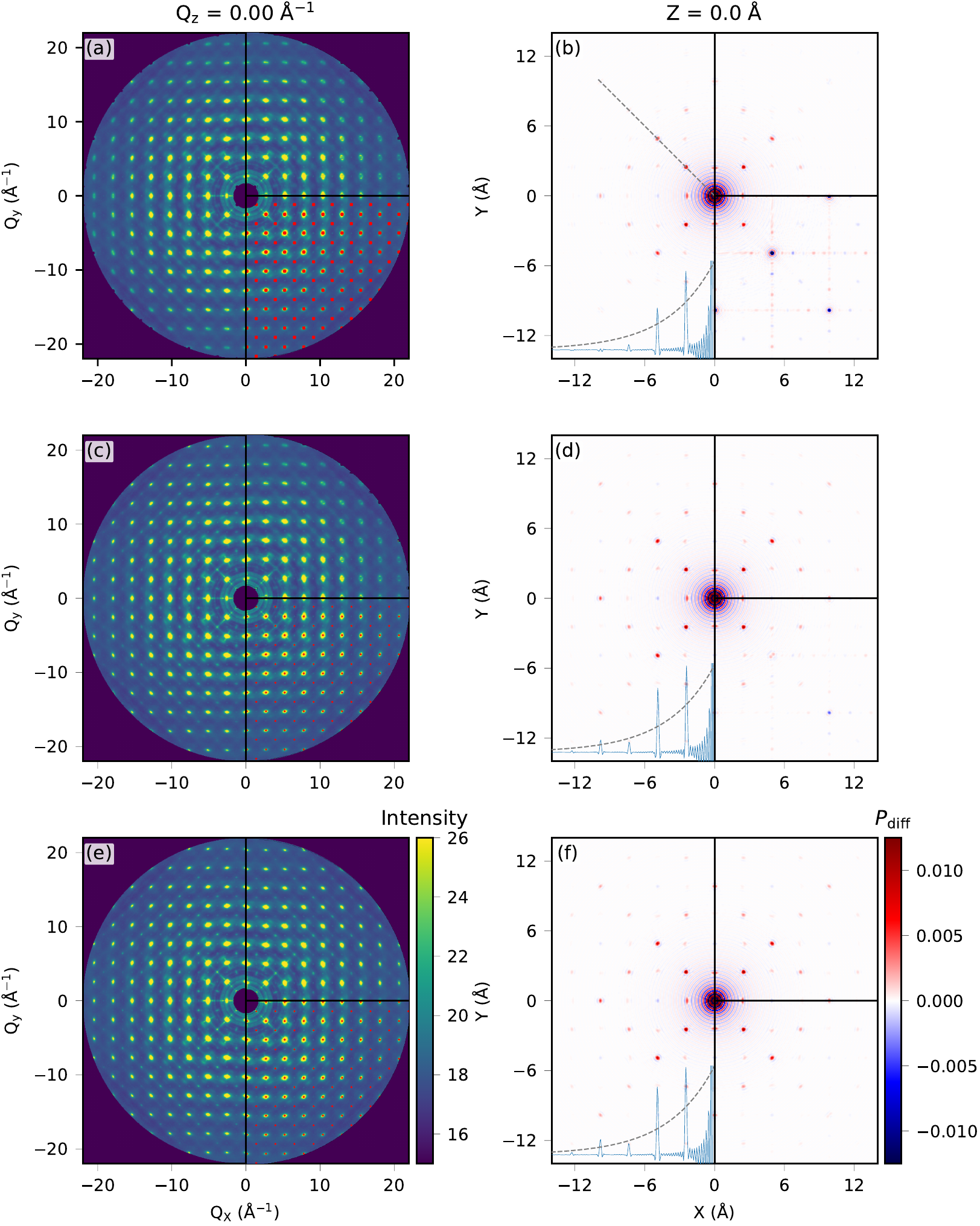}
\caption{The choice of Bragg peak punch size:
Representative slices of reciprocal space and associated slices of the 3D-$\Delta$PDF after punching and filling with a punch size of (a,b) 9 voxels, (c,d) 5 voxels, and (e,f) 3 voxels.
The lower right quadrants of (a, c, e) show the punch used in red, while the upper right quadrants show the result of a linear interpolation filling \rjkadd{and the left halves show the result of a DCT interpolation filling.}
The lower right quadrants of (b, d, f) show the 3D-$\Delta$PDF result of filling punched areas uniformly with zero, while the upper right quadrants show the result of a linear interpolation filling \rjkadd{and the left halves show the result of a DCT interpolation filling.}
\rjkadd{Inset in the lower left corners of (b, d, f) are 3D-$\Delta$PDF profiles along the $\langle 110\rangle$ direction depicted by the dashed straight line in (b).
Each inset is plotted on an identical scale, and contains an identical but arbitrary exponential decay curve for comparison.
Differences observed across punch sizes and/or filling technique are discussed in the text.}
\soutoldold{The differences are nearly impossible to detect visually within the reciprocal space maps, even when using linear interpolation for filling.
With the exception of the most extreme case, a very large punch filled with zero intensity, only very small differences are apparent between the 3D-$\Delta$PDFs, primarily within the absolute intensity of the $\Delta$PDF peaks.
Qualitatively, the 3D-$\Delta$PDF are nearly identical.}}
\label{fig:punch_size}
\end{figure}

The size of the voxel punch is relevant, as a punch which is too large can remove important features in close proximity to Bragg peaks, \rjkadd{ often corresponding to long-range features in direct space}, while a punch which is too small can leave behind Bragg intensity tails.
In our work, we have tested three different punch sizes, with a diameter of either 3, 5, or 9 voxels, spanning 0.189~\RAA, 0.315~\RAA, or 0.567~\RAA, respectively.
For each reciprocal space data-set, we have computed the location within the associated data array of all Bragg peaks based on both the known crystal unit cell size/shape and the associated data array $Q$-step.
We have applied a punch at this location, which is computationally handled by setting array values inside the punch to ``not a number" (NaN) to mark the locations for subsequent filling.
Examples of the punch effect in a line scan of intensity in reciprocal space are shown in Fig.~\ref{fig:punch_line}(a), while similar examples for a slice of reciprocal space intensity are shown in the lower right quadrants of Fig.~\ref{fig:punch_size}(a, c, e).
It is clear in Fig.~\ref{fig:punch_line}(a) that each punch, even the smallest tested, effectively removes the large Bragg peaks from the data.
This is reasonable, if we recall that that the Bragg peaks have a maximum full-width at half-maximum of about 0.19~\RAA\ (see e.g. Fig.~\ref{fig:bragg_peaks}).

Once all Bragg peaks have been punched from the data, a suitable filling or interpolation algorithm must be chosen so as to replace any diffuse intensity co-located with the punched Bragg peaks.
In the case considered here, where we have elected to operate on the reciprocal space intensity distribution, our filling algorithm must be capable of interpolating a three-dimensional function.
In situations where the diffuse intensity surrounding and underneath the Bragg peaks is relatively flat or \esbaddold{\soutoldold{diffuse}broad}, simple linear or quadratic interpolation may be sufficient.
It is clear however that the diffuse intensity distribution within the data shown in Fig.~\ref{fig:punch_line}(a) contains peak-like features centered at the location of Bragg peaks, although the diffuse peaks are significantly broader and an order of magnitude less intense (see e.g. Fig.~\ref{fig:symm}).
In this situation, linear interpolation may be insufficient.

\esbaddold{\soutoldold{In this work}Here}, we have tested two simple filling techniques and a third more complex technique.
For the simpler approaches, punched portions of reciprocal space were filled either with zero intensity or by using linear interpolation.
As a more complex approach, we have adopted an algorithm typically used when handling large three-dimensional geophysical data-sets~\cite{wang_three-dimensional_2012}.
This approach \rjkadd{fills gaps by iteratively updating a reconstructed, gap-less data-set with weighted residuals propagated through inverse and forward} discrete cosine transforms (DCTs).
\rjkadd{It} has been found to produce a global normalized error of less than $5\times10^{-5}$ in test data-sets ~\cite{garcia_robust_2010}.

To investigate the result of our punch and fill tests, we again look at line scans of the reciprocal space intensity distribution under a number of situations, shown in Fig.~\ref{fig:punch_line}.
The result of filling this punched data with linear interpolation are shown in Fig.~\ref{fig:punch_line}(b), while the use of an iterative DCT interpolation algorithm are shown in Fig.~\ref{fig:punch_line}(c).
The trivial case of zero-filling is not shown.

As expected, linear interpolation creates clear discontinuities in the slope of the filled intensity distribution, apparent in intensity line scans (Fig.~\ref{fig:punch_line}(b)).
These discontinuities are less apparent in the maps of reciprocal space intensity shown in the upper right quadrants of Fig.~\ref{fig:punch_size}(a, c, e), likely because they are obscured by the narrow color scale, chosen to highlight weak features.

Conversely, the DCT filling routine largely preserves the peak-like feature underneath the Bragg peak upon filling, as can be seen in Fig.~\ref{fig:punch_line}(c).
The relative intensities of these diffuse peaks are also preserved, with the largest difference being within the absolute intensities of this peak.
We do note that filling of the largest 9-voxel punch does lead to some anomalous 'crater' like features, suggesting that this punch is perhaps too large, or that the iterative DCT filling routine has failed to converge in some situations.

Ultimately, the most important aspect of the punch and fill process is the impact on the full 3D-$\Delta$PDF itself, and as such we will visit this topic here even if the Fourier transform processing step will not be introduced until the next section.
The result of varying the punch size and filling approach is shown in representative reciprocal space intensity maps and the associated 3D-$\Delta$PDFs in Fig.~\ref{fig:punch_size}.
\rjkadd{Also shown in Fig.~\ref{fig:punch_size}(b, d, f) for quantitative comparison are line scans along the $\langle 110\rangle$ direction depicted by the dashed straight line in panel (b). 
Each line scan is superimposed with an identical exponential decay curve (dashed line) so that the relative heights of the 3D-$\Delta$PDF features can be compared.}

Clearly in our example, the choice of punch size has \rjkadd{\soutold{no}little} qualitative impact on the final 3D-$\Delta$PDFs when combined with an iterative DCT filling routine, as each contains the same key features.
\rjkadd{\soutold{Subtle differences do exist in the absolute intensity of these features, specifically, as the punch size is reduced, the peaks within the 3D-$\Delta$PDFs become more intense.}}
\rjkadd{The line scans shown inset in Fig.~\ref{fig:punch_size}(b, d, f) reveal that there are subtle quantitative differences in the relative intensities of the 3D-$\Delta$PDF features when moving from the largest (9-voxel) punch to the intermediate (5-voxel) punch.
Specifically, features decay more quickly as a function of pair distance in the the 3D-$\Delta$PDF associated with the 9-voxel punch.
Coversely, the feature decay rate is preserved when comparing the intermediate (5-voxel) and smallest (3-voxel) punch.
With the exception of the small feature at 10~\AA\ along the $\langle 110\rangle$, the relative intensities of these features are otherwise largely unchanged.}

\rjkadd{It is then important to note the aim of the 3D-$\Delta$PDF study prior to setting out on measurement and data reduction.
It is possible that a qualitative appraisal, based on e.g. the presence/absence of certain features, or the relative signs of two or more related features~\cite{schaub_exploring_2007}, is sufficient, and in this case, the 3D-$\Delta$PDF may not be particularly sensitive to punch size and filling algorithm.
It is also possible that detailed quantitative analysis is required, where the relative intensity and/or decay rate of features as a function of pair distance is critically important~\cite{holm_temperature_2020}.
\rjkadd{For such analysis it is clear that the choice of punch size is relevant, especially in cases where diffuse features are peak-like and co-located with Bragg peaks in reciprocal space.
A best-practice would be to begin with a punch larger than necessary, and to gradually reduce the size until features in the 3D-$\Delta$PDF stop changing.
Additionally, one can investigate the obtained 3D-$\Delta$PDF to be sure that the features obtained when using a punch of a given size are physically consistent with the average atomic structure.}
One must adjust the measurement and reduction details such that the overhead matches the level of detail required in the final analysis, and be prepared to conduct further measurements when a disparity occurs.}

\section{Fourier Transform}
\label{section:ft}

\begin{figure}
\includegraphics[width=0.8\textwidth]{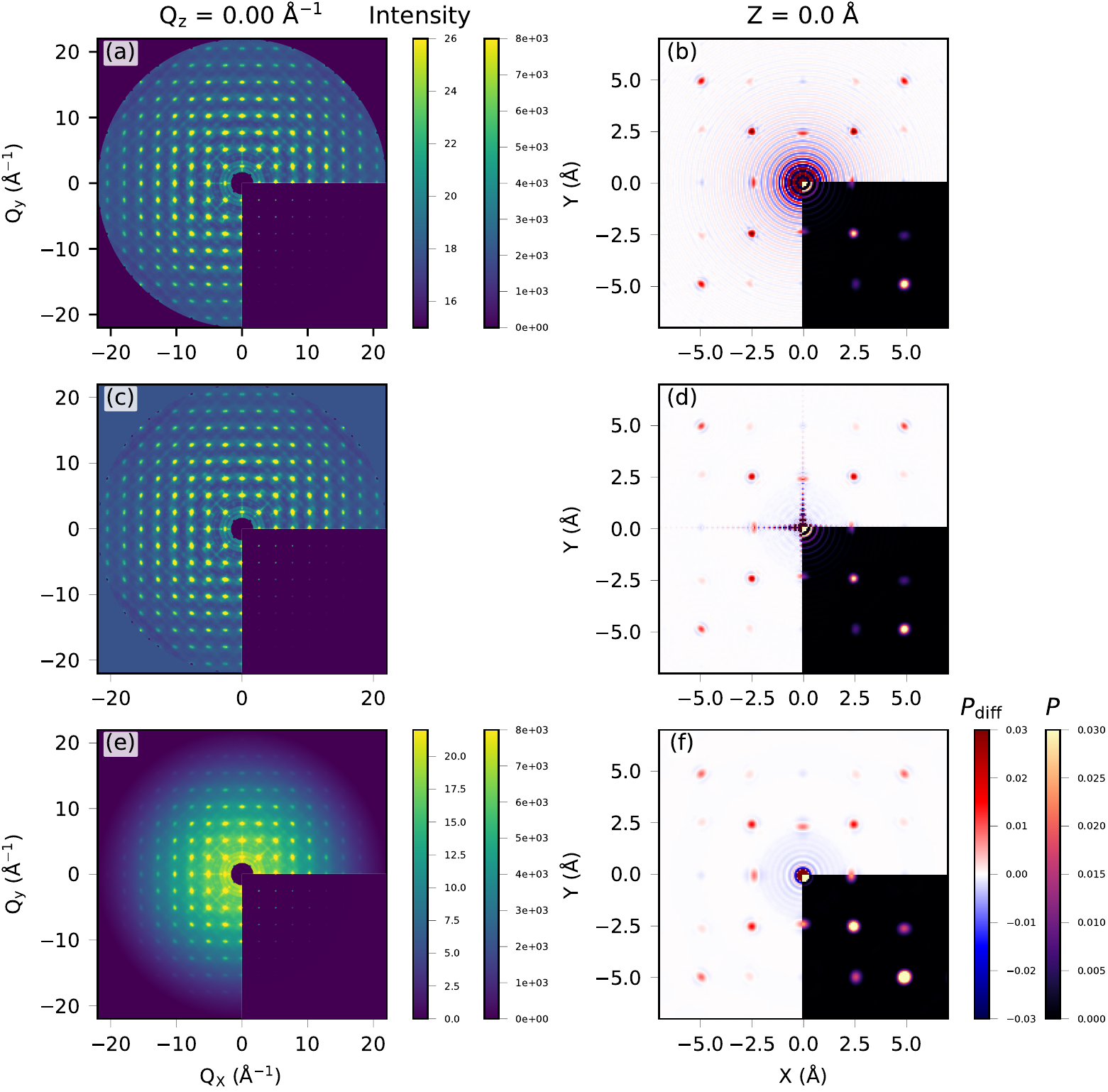}
\caption{Application of a damping window function:
\firstrevadd{\soutoldoldold{Representative slices of reciprocal space and an associated slice of the 3D-$\Delta$PDF (a,b) before and (c,d) after the application of the damping window function as described in the text.
The lower right quadrants of (a,b) and (c,d) show the result of retaining all Bragg peaks in reciprocal space and the associated full 3D-PDF, respectively.
The primary impact is the reduction of Fourier transform termination ripples as can be seen in panel (d).}
Representative slices of reciprocal space and an associated slice of the 3D-$\Delta$PDF (a,b) as processed, (c,d) after filling outside \qmax with a constant value as described in the text, and (e,f) after the application of the damping window function as described in the text.
The lower right quadrants of (a,c,e) and (b,d,e) show the result of retaining all Bragg peaks in reciprocal space and the associated full 3D-PDF, respectively.
The primary impact is the reduction of Fourier transform termination ripples as can be seen in panel (f).
}}
\label{fig:ft_taper}
\end{figure}

Once a satisfactory reciprocal space intensity distribution has been obtained and the Bragg intensity removed, obtaining the full 3D-$\Delta$PDF requires applying a \esbaddold{discrete Fourier transform (DFT)}.
As the intensity distribution is a real valued even function (it obeys inversion symmetry) its Fourier transform is also a real valued even function, and as such only the real part of the Fourier transform is \rjkadd{non-zero\soutold{required}}.

\subsection{Window Function}
\label{section:window}
Following detector to reciprocal space remapping, data merging, and the application of all symmetry operations, the filled portion of reciprocal space is often irregular.
In order to \esbaddold{\soutoldold{prevent}avoid appearance of} problematic Fourier effects, it is \esbaddold{\soutoldold{useful}helpful} to apply a window function to the reciprocal space intensity distribution prior to applying the DFT.
Many options for this exist, including filling the area outside a given radius with a constant intensity value~\cite{roth_solving_2019,krogstad_reciprocal_2020}.
Choosing this intensity value can be problematic however, and can cause discontinuities in the reciprocal space intensity which will manifest \esbaddold{as artifacts} in the 3D-$\Delta$PDF.
An extreme case of this is shown in Fig.~\ref{fig:ft_taper}(a), where data outside a scattering vector magnitude of 21~\RAA\ have been set to zero, effectively representing a hard sphere window function.
In Fig.~\ref{fig:ft_taper}(b) we show the resulting 3D-$\Delta$PDF from this hard sphere window function, which is heavily impacted by the window function ripples\esbaddold{, observable as concentric circles in the figure,} superimposed over the origin.
These ripples propagate far into the 3D-$\Delta$PDF, effectively corrupting the features nearest and second nearest the origin.
\firstrevadd{Similar plots resulting from filling outside 21~\RAA\ with non-zero values, selected as the median intensity between 20.5 and 21~\RAA, are also shown in Fig.~\ref{fig:ft_taper}(c,d).
This can remedy the Fourier ripples somewhat, but selecting this non-zero constant filling is rather arbitrary.}

The alternative adopted here is the application of, \firstrevadd{through point-wise multiplication}, a modified Lorch function $L_1$ of the formula~\cite{soper_use_2012}
\begin{equation} \label{eq:lorch}
L_1(Q,\Delta_1) = [3/(Q\Delta_1)^3](\sin Q\Delta_1 -Q\Delta_1 \cos Q\Delta_1)
\end{equation}
where $Q$ is the magnitude of the momentum transfer scalar and $\Delta_1$ is the smearing radius in reciprocal space, here taken as 21~\RAA.
\esbaddold{\soutoldold{This }A similar approach} is \esbaddold{\soutoldold{commonly}sometimes} used in the powder total scattering community when computing the full 1D-PDF\oldrobadd{~\cite{lorch_neutron_1969,soper_extracting_2011}}.
An example of the application of this damping window function to our reciprocal space intensity distribution is shown in Fig.~\ref{fig:ft_taper}(e).
Here, it can be seen that the intensity decreases uniformly towards zero at the boundary of reciprocal space.
The 3D-$\Delta$PDF resulting from the application of this \esbaddold{\soutoldold{smoothed}damping} window function is shown in Fig.~\ref{fig:ft_taper}(f).
Notably, the ripples present in Fig.~\ref{fig:ft_taper}(b) have been effectively mitigated, yielding a clearer picture of the features nearest and second nearest the origin.
Although it is known that the application of such a \esbaddold{\soutoldold{smoothed}damping} window function does reduce the effective resolution of the data, we can see from Fig.~\ref{fig:ft_taper}(f) that even after its application, we can still resolve each individual 3D-$\Delta$PDF feature.
\firstrevadd{If fine resolution may be required, one should naturally carefully compare the filtered and unfiltered 3D-$\Delta$PDF}.
%
%
%
%
\section{Robustness}
\label{section:robustness}
Given that this work has taken great length to expand upon many of the steps in obtaining the 3D-$\Delta$PDF, it is worth questioning the robustness of this approach, not only against errors or omissions in data collection and processing, but also against a variation in observations between physically unique samples of a given crystal.

\subsection{Data Processing Sequence}
\label{section:sequence}
\begin{figure}
\includegraphics[width=0.8\textwidth]{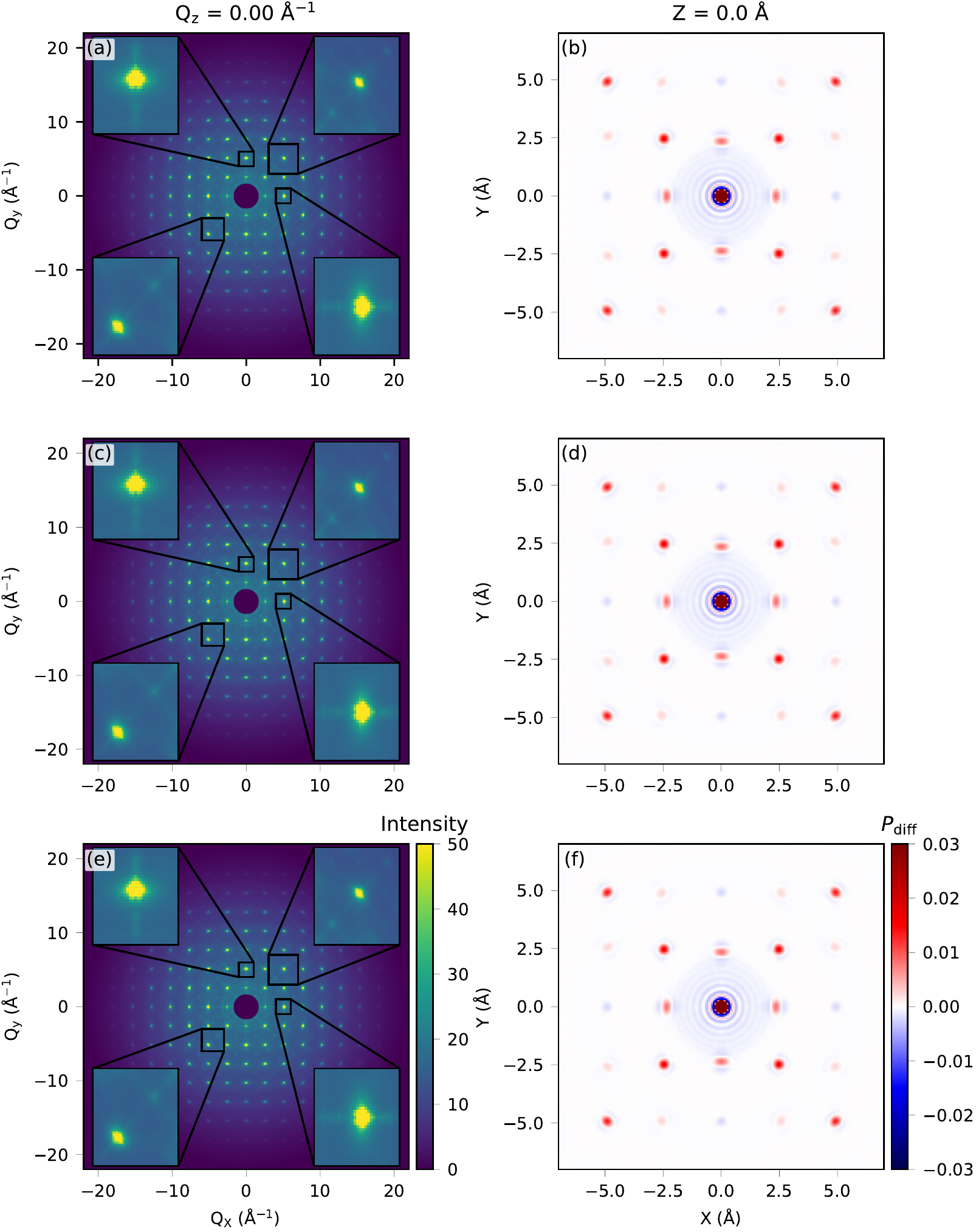}
\caption{The role of work-flow ordering on the final intensity distribution and 3D-$\Delta$PDF:
Representative slices of reciprocal space and an associated slice of the 3D-$\Delta$PDF when varying the ordering of the data processing steps.
\rjkadd{A few regions of interest in reciprocal space are shown inset on an enlarged scale.}
(a, b) Data were symmetrized, different exposure times were merged, punched, filled, and then Fourier transformed.
(c, d) Data were symmetrized, punched, filled, Fourier transformed, and then different exposure times were merged.
(e, f) Data were punched, filled, symmetrized, different exposure times were merged, and then Fourier transformed.
Visually the difference is nearly impossible to detect within the reciprocal space maps.}
\label{fig:proc_order}
\end{figure}
In general, raw detector images which have been dynamically masked to mitigate the effects of detector afterglow and blooming, background subtracted, corrected for \rjkadd{\soutold{absorption}interframe scale fluctuation} effects, and transformed to crystal reciprocal space can subsequently be processed in any number of ways.
With the only requirement that punching and filling of Bragg peaks occurs prior to DFT, \esbaddold{\soutoldold{we are free to}} merging multiple data-sets, applying symmetry operations, and removing Bragg intensity \esbaddold{can be done} in any order.
The merging and the application of \rjkadd{weighted} symmetry averaging can also be applied to the 3D-$\Delta$PDF itself, as it should adhere to the same point symmetry as the reciprocal space intensity distribution.

Given that the choice of work-flow ordering is largely arbitrary, it is interesting to explore the extent to which the 3D-$\Delta$PDF is reproducible across different orderings of \esbaddold{\soutoldold{our}the} data processing \esbaddold{steps}.
To investigate this, we have tested all possible permutations of data processing steps, each with the three distinct Bragg peak punch sizes outlined in section~\ref{section:punchfill}.
The reciprocal space intensity distributions and 3D-$\Delta$PDFs resulting from a subset of these permutations are shown in Fig.~\ref{fig:proc_order}.
Remarkably, both the reciprocal space intensity distributions and 3D-$\Delta$PDFs are quite robust against variations in work-flow ordering, being qualitatively identical.
\firstrevadd{Quantitatively, these reciprocal space intensity distributions show an average and maximum $R_{\textrm{diff}}$ of 0.2~\% and 1.0~\%, respectively.}
While this may not be surprising, it is encouraging to note that the final 3D-$\Delta$PDF is largely invariant under these conditions.

\subsection{Crystal Variation and Mounting}
\label{section:mount}
\begin{figure}
\includegraphics[width=0.8\textwidth]{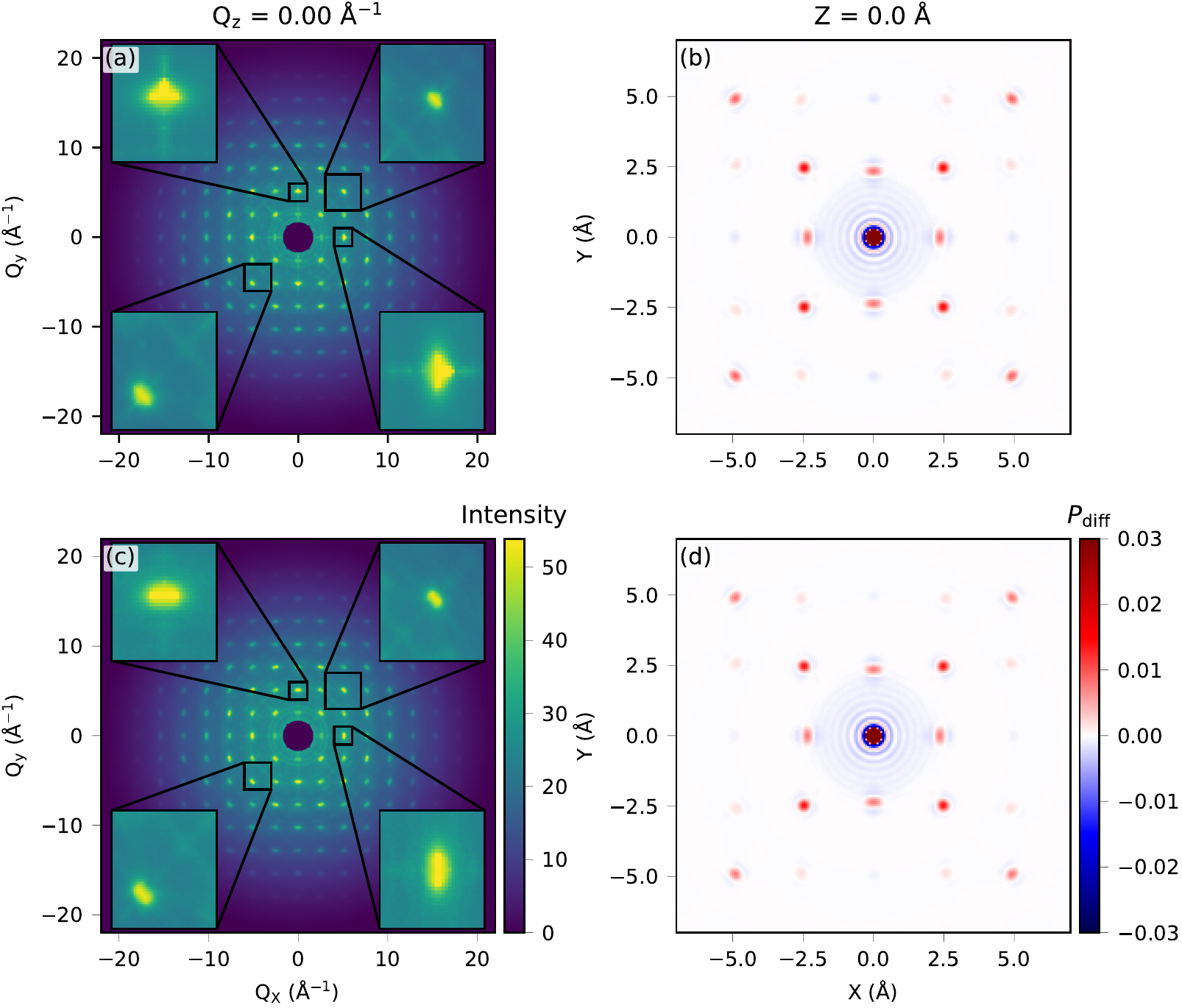}
\caption{Sample-to-sample reproducibility in the final intensity distribution and 3D-$\Delta$PDF:
Representative slices of reciprocal space and an associated slice of the 3D-$\Delta$PDF measured and processed from distinct \cis\ crystals.
\rjkadd{A few regions of interest in reciprocal space are shown inset on an enlarged scale.}
(a,b) Data collected and processed from the single crystal of \cis, \rjkadd{sample 1,} featured throughout the manuscript.
(c,d) Data collected and processed from\rjkadd{,sample 2,} a different single crystal of \cis.
Both the reciprocal space maps and the 3D-$\Delta$PDFs are nearly identical, demonstrating the reproducibility of the technique.}
\label{fig:sample_to_sample}
\end{figure}

3D-$\Delta$PDF analysis relies on detection of quite subtle features within the reciprocal space intensity distribution.
Given this, it is interesting to explore the extent to which these subtle features are reproducible between repeated measurements of distinct crystals of the same material.
To address this \esbaddold{\soutoldold{concern}matter}, we have measured the full reciprocal space intensity distribution from two distinct \cis\ crystals.
These samples are of comparable quality and show identical physical properties.
\esbaddold{Their primary difference from the experimental standpoint are slightly different} physical dimensions and a distinct crystal orientation dictated by the random mounting process.
A representative reciprocal space map and the 3D-$\Delta$PDF from each sample are shown in Fig.~\ref{fig:sample_to_sample}.
When comparing the two samples, we note \rjkadd{only slight variations in the persistence of streaking in the reciprocal space intensity distributions, visible within the insets of Fig.~\ref{fig:sample_to_sample}(a, c).
This is likely due to variations in the effectiveness of our dynamic masking heuristic between the two physically distinct crystals.} 
\firstrevadd{Quantitatively, the two intensity distributions show $R_{\textrm{diff}}= 0.1~\%$, comparable to the mean value obtained by varying the order of data reduction steps.} 
\rjkadd{Despite these small variations in reciprocal space, we observe nearly no differences in} direct space.
This demonstrates that the technique is quite robust against sample to sample variation imposed by crystal size or mounting.
\firstrevadd{\soutoldoldold{It is worth noting that crystal orientation as it pertains to mounting may become relevant when dealing with low symmetry systems and/or a limited rotation angle during measurement.
In these cases, care must be taken to orient and rotate the crystal so as to cover all of reciprocal space within the desired $Q$-range.}}

\subsection{\rjkadd{\soutold{Minimally}Variably} Processed Data}
\label{section:minproc}

Our results thus far suggest that both the reciprocal space intensity distribution and the 3D-$\Delta$PDF measured from a given material are robustly reproducible in a number of situations.
As a final test, we envision \rjkadd{\soutold{sub-optimal}several hypothetical} scenarios for both measurement and data processing: a single data-set with minimal counting of 0.1~s per frame, with no \rjkadd{\soutold{absorption}interframe scale} correction or background subtraction, where outliers have been included in all steps, and a small (3-voxel) Bragg punch has been applied.
\oldrobadd{We have tested \rjkadd{\soutold{two}three} permutations of this processing.
First, we have used only linear filling at the location of punched Bragg intensity, and omitted both the relevant symmetry operations and any dynamic masking.
This represents the most pessimistic case.
\rjkadd{Second, we have used DCT filling of punched Bragg intensity, and included the relevant symmetry operations, but omitted any dynamic masking.}
\rjkadd{\soutold{Second}Third}, we have used DCT filling, have included dynamic masking, and applied the relevant symmetry operations so as to ensure the reciprocal space intensity distribution fully fills our voxel map.}

\begin{figure}
\includegraphics[width=0.75\textwidth]{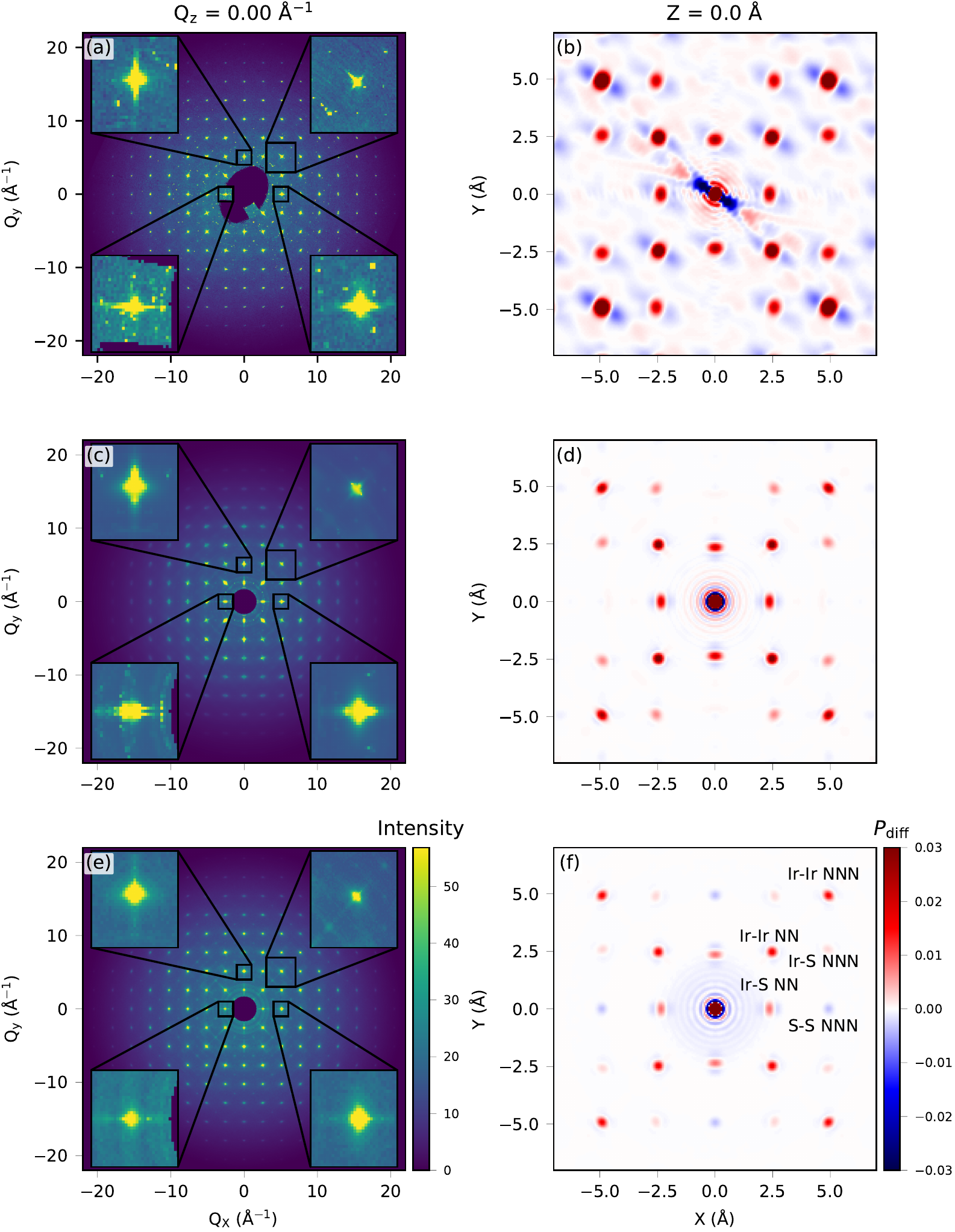}
\caption{\rjkadd{\soutold{Minimally} Variably} processed data-sets:
(a, c\rjkadd{, e}) Representative slices of reciprocal space and (b, d\rjkadd{, f}) the associated slices of the 3D-$\Delta$PDF which have been taken through fewer data processing steps.
\rjkadd{A few regions of interest in reciprocal space are shown inset on an enlarged scale to emphasize the differences associated with each level of processing.}
\oldrobadd{Data in (a,b) represent one single full crystal rotation, with an exposure time of just 0.1~s per frame, where both dynamic masking and symmetry averaging have been omitted.
\rjkadd{Data in (c, d) have undergone the full extent of data processing (merging exposure times, symmetry averaging), but have not undergone dynamic masking. 
A 3-voxel punch and DCT fill process was used, with a Lorch window function prior to FT.}
Data in \rjkadd{\soutold{(c, d)}(e, f)} represent one single full crystal rotation, with an exposure time of 0.1~s per frame.
In this case, data have been remapped to reciprocal space after dynamically masking, have undergone symmetry averaging and a 3-voxel punch and linear fill process, with a Lorch window function prior to FT.}
In all cases, no background subtraction or \rjkadd{\soutold{absorption}interframe scale} corrections have been applied.
\rjkadd{Observations regarding these different processing pathways are discussed in the text.}
\soutoldold{Panels (a,b) are severely compromised, with substantial issues present in both the intensity distribution and 3D-$\Delta$PDF}
\soutoldold{Remarkably, the 3D-$\Delta$PDF in (d) is nearly indistinguishable from those subject to additional processing (see e.g. Fig.~\ref{fig:sample_to_sample}),
Panel (d) also contains feature labels for different known pair correlations in the \cis\ system.}
}
\label{fig:min_proc}
\end{figure}

Representative slices of the reciprocal space intensity distributions and the associated 3D-$\Delta$PDFs are shown in Fig.~\ref{fig:min_proc}.
In the worst case, \rjkadd{shown in panels (a, b)} with linear filling and without symmetrization or dynamic masking, both the reciprocal space intensity distributions and 3D-$\Delta$PDFs show spurious features.
The lack of symmetry averaging has produced a reciprocal space intensity distribution which does not uniformly fill $\mathbf{Q}$-space.
\rjkadd{It is clear from the insets in Fig.~\ref{fig:min_proc}(a) that, without dynamic masking, there is substantial streaking still present in the diffracted intensity distribution.
Additionally, ring-like collections of parasitic Bragg peaks, previously removed by our dynamic masking, are again now visible near the origin in Fig.~\ref{fig:min_proc}(a).}
\firstrevadd{Compared to a fully processed data-set, the intensity distribution in Fig.~\ref{fig:min_proc}(a) gives an excessively large $R_{\textrm{diff}}= 143~\%$.} 

The 3D-$\Delta$PDF \rjkadd{associated with this intensity distribution, shown in Fig.~\ref{fig:min_proc}(b), is characterized by} strong unphysical artifacts with the same intensity scale as the relevant atom-pair features.
This is likely due to both the incomplete nature of the reciprocal space data-set and also the spurious features not removed by dynamic masking.
This scenario is only marginally similar to those shown in e.g. Fig.~\ref{fig:sample_to_sample}, suggesting \esbaddold{that this} would not be useful even for a cursory screening of the presence of local distortions.

The reciprocal space intensity distribution and associated 3D-$\Delta$PDF shown in Fig.~\ref{fig:min_proc}(c, d) represent a significant improvement over those shown in Fig.~\ref{fig:min_proc}(a, b).
These data were processed similarly, with symmetry averaging included, and with DCT rather than linear filling.
In both cases, dynamic masking was excluded.
Notably, the inclusion of symmetry averaging has produced a reciprocal space intensity distribution which does uniformly fill $\mathbf{Q}$-space.
It has also mitigated some spurious features, which can be seen when comparing the upper left insets of Fig.~\ref{fig:min_proc}(a) and (c).
Unfortunately, as dynamic masking was not included, streaking is still present, and symmetry averaging has actually compounded the issue.
This is apparent when comparing the upper right or lower left insets of Fig.~\ref{fig:min_proc}(a) and (c), where streaks have been clearly propagated by symmetry averaging.
Additionally, ring like features are actually exacerbated by symmetry averaging.
\firstrevadd{Compared to a fully processed data-set, the intensity distribution in Fig.~\ref{fig:min_proc}(c) gives $R_{\textrm{diff}}= 89~\%$, a significant improvement over that shown in Fig.~\ref{fig:min_proc}(a), but still rather large.} 

The 3D-$\Delta$PDF associated with this intensity distribution is shown in Fig.~\ref{fig:min_proc}(d).
The inclusion of symmetry averaging has substantially improved the appearance of the 3D-$\Delta$PDF when compared to that in panel (b).
Indeed, when compared to the 3D-$\Delta$PDFs shown in e.g. Fig.~\ref{fig:sample_to_sample}, both contain the same features, and the general properties of each features are retained. 
Some spurious features however do remain, with significant quantitative differences, and there appear to be ring-like features superimposed, likely due to the ring-like features in the reciprocal space intensity distributions.

\rjkadd{Fig.~\ref{fig:min_proc}(e, f) shows the results when dynamic masking is added to the processing steps that produced the data shown in Fig.~\ref{fig:min_proc}(c, d).
The insets in panel (e) suggest that the inclusion of dynamic masking when transforming from detector to reciprocal space (prior to merging or symmetry averaging) significantly reduces the streaking and parasitic ring-like scattering seen in Fig.~\ref{fig:min_proc}(a, c).}
Remarkably, \oldrobadd{in this case, the 3D-$\Delta$PDF shows intense, well resolved features.
Indeed,} the reciprocal space intensity distributions and 3D-$\Delta$PDFs for the \rjkadd{case including dynamic masking} in Fig.~\ref{fig:min_proc}(e, f) are both qualitatively identical to those shown in e.g. Fig.~\ref{fig:sample_to_sample}(a, b), where quite extensive data processing has been conducted.
\firstrevadd{Although qualitatively identical,  the intensity distribution in Fig.~\ref{fig:min_proc}(e) gives $R_{\textrm{diff}}= 54~\%$ when compared to to a fully processed data-set.
This represents a significant improvement over that shown in either Fig.~\ref{fig:min_proc}(a) or (c), but may be too large for useful quantitative analysis.} 

Nonetheless, with an exposure time of just 0.1~s per frame and a rotation step size of 0.1\ignorespaces\textdegree\, the data shown in Fig.~\ref{fig:min_proc} were collected in about 6 minutes.
This result implies that the \rjkadd{qualitative features present in the} 3D-$\Delta$PDF can be reproducible even under the non-ideal conditions and with marginal data processing,
\oldrobadd{and that, while all the data processing steps discussed herein contribute to a clean 3D-$\Delta$PDF, some steps, such as dynamic detector masking, are more essential than others.
Of course, it is always the aim to collect as high-quality data as possible, but this can unfortunately come at the \esbaddold{\soutoldold{cost of resources}expense of efficient resource utilization} which could be more effectively spent probing additional states of the system (e.g. temperature dependence) or even other systems.
It is thus advantageous to know the point at which diminishing returns are achieved when collecting and/or processing 3D-$\Delta$PDF data.}

\oldrobadd{It is important to point out here that these pessimistic scenarios were partially successful in the \cis\ case, but that more complex systems may present additional challenges, requiring full data-processing to garner \rjkadd{\soutold{useful results}usable data}.
Related to this, the use of different 2D detectors may yield different results, especially if the detector benefits from additional dynamic range.
That being said, even large improvements in dynamic range are unlikely to eliminate all detector artifacts, and some improved detectors create new issues.
For example, spatial module gaps in the PILATUS series of detectors~\cite{broennimann_pilatus_2006,kraft_pilatus_2010} create further holes in reciprocal space, where symmetrization, merging of distinct data-sets with slightly different detector positions~\cite{roth_solving_2019,davenport_fragile_2019, holm_temperature_2020, krogstad_reciprocal_2020}, \rjkadd{and/or crystal rotation around multiple axes} are \rjkadd{\soutold{expected to be}} of increased importance.
\firstrevadd{A blooming-like effects has also been reported in these more modern detectors~\cite{krogstad_reciprocal_2020}, underscoring the need for a continuing discussion on measurement artifacts.}
}

\begin{figure}
\includegraphics[width=0.8\textwidth]{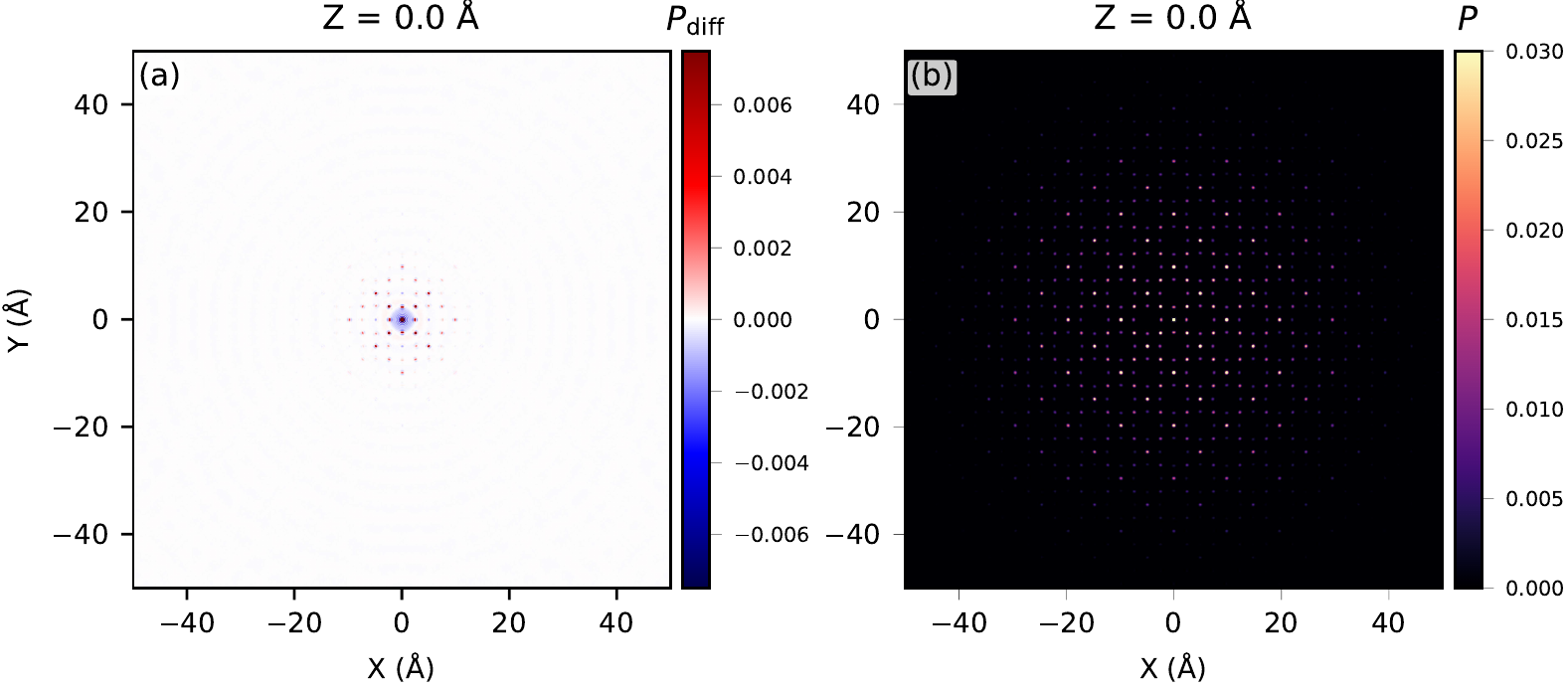}
\caption{\oldrobadd{3D-PDF and 3D-$\Delta$PDF \rjkadd{\soutold{results}data}:
(a) A representative fully-processed slice of the 3D-$\Delta$PDF\rjkadd{, where DCT filling and a 5-voxel punch was used}.
Note that the majority of signal decays to zero by $r=10$~\AA, suggesting that any deviation between the local and long-range range structure has a finite correlation length in direct-space.
In the \cis\ system, 10~\AA\ corresponds to about one unit-cell.
(b) A representative fully-processed slice of the full 3D-PDF, obtained by retaining the full reciprocal space intensity distribution (not removing Bragg peaks).
Note that the majority of signal decays to zero by $r=35$~\AA, highlighting the effect of finite $\mathbf{Q}$-space resolution.}
\esbaddold{To emphasize weak features, color scales in both panels are adjusted such that the strong features at shorter interatomic vectors are saturated.}
}
\label{fig:corr_length}
\end{figure}

\subsection{Material Specific Observations}
\label{section:cisspecific}

Finally, we explore some \rjkadd{qualitative} observations \rjkadd{regarding the 3D-$\Delta$PDF} that are specific to \cis\ system. 
\rjkadd{We note that a full quantitative analysis will be the subject of subsequent work, and requires a full description of any and all disorder present in the long range average structure to which the 3D-$\Delta$PDF is referenced, including e.g. anisotropic atomic displacement parameters and atomic coordinates/occupancies. 
Care must be taken to check that any interpretation is consistent with this full description of the long range average structure, as the 3D-$\Delta$PDF can be misinterpreted in some cases.
The following is based on an assumed, minimally disordered, literature published description of the long range average structure~\cite{bozin_local_2019}.}
In Fig.~\ref{fig:min_proc}(f) we label a few selected interatomic vector contributions for the $Z=0$ cut of the 3D-$\Delta$PDF.
\rjkadd{This labeling is not exhaustive, and pairs which do not demonstrate obvious connectivity (i.e. Cu-Cu nearest neighbor (NN) pairs which exist in separate CuS$_4$ tetrahedra) are not labeled.} 
The strongest observed contributions to the differential \oldrobadd{\soutoldold{are coming from} are due to} Ir-Ir NN and next-nearest neighbor (NNN) pairs \oldrobadd{along the $\langle 110\rangle$ family of directions}.
This is not surprising, as \oldrobadd{Ir is the strongest scatterer in the system and }the nature of the local distortion involves predominantly \oldrobadd{the} Ir sublattice~\cite{bozin_local_2019}\rjkadd{, where strong bonding is present}\soutoldold{, and since Ir is the strongest scatterer in the system}.
Importantly, the same cut also shows observable, albeit weaker, Ir-S NNN and S-S NNN contributions, as labeled.  
\rjkadd{The S-S NNN contributions are themselves unusual, as they appear as faint peaks of apparent negative intensity.
This is strictly a function of the color scale, which was chosen to emphasize stronger features.
The fully three-dimensional S-S NNN feature consists of a negative central lobe surrounded by a weaker diffuse positive outer lobe, suggesting it is associated with some sort of negatively correlated displacive disorder~\cite{weber_three-dimensional_2012}. 
The integral of this feature over a cube (1.5~\AA\ on a side) is zero within a small tolerance.}

The significance of \rjkadd{\soutold{this, particularly of} a} non-zero differential intensity corresponding to the NNN S-S pair, is as follows. 
\rjkadd{In the presence of a similar pair-correlation strength for the underlying local structural distortion,} the S-S intensity is expected to be $\sim$ 20 times weaker than the Ir-Ir contribution, since \rjkadd{the two pairs have identical multiplicity, and} for a given pair of atoms the PDF intensity scales with a product of their scattering form factors.
In 1D-PDF of \cis, where the three-dimensional information is lost due to powder averaging, any S-S NNN information \oldrobadd{associated with a local structural distortion} is effectively suppressed not only because S scatters x-rays much more weakly than Ir does, but also due to significant overlap of S-S NNN with Ir-Ir NN PDF peaks~\cite{bozin_local_2019}.
\rjkadd{Within the 3D case, Ir-Ir NN and S-S NNN contributions do not overlap, as the corresponding interatomic vectors are in different directions, as can be seen in Fig.~\ref{fig:sample_to_sample}(f).}

\oldrobadd{It is also important to note that within 1D-PDF analysis, Bragg information becomes mixed with any local structural signal, and disentangling the two is non-trivial during structural fitting, where phenomenological modeling of correlated atomic motion can effectively remedy what is in actuality a local structural distortion\cite{jeong;jpc99,jeong;prb03}.}
The ability to \oldrobadd{directly} observe \oldrobadd{a} differential S-S NNN signal in the 3D-$\Delta$PDF of \cis\ therefore implies that\soutoldold{, since different atomic pair contributions are resolved in 3D,} the sensitivity of the 3D-$\Delta$PDF approach to local deviations in systems comprised of light and heavy elements is not limited to heavier atoms, but is also suitable\rjkadd{, at least in principle,} for exploring local correlations among weaker scatterers.

\esbaddold{The spatial extent of local distortions is considered in Fig.~\ref{fig:corr_length}(a), where we plot the $Z=0$ slice of \cis\ 3D-$\Delta$PDF, containing Ir-Ir contributions, over an extended spatial range.
From this one sees that the extent of local structural correlations is limited to sub-nanometer lengthscale, in agreement with powder measurements~\cite{bozin_local_2019}. 
The ability to assess the extent of local structural correlations \oldrobadd{directly} using 3D-$\Delta$PDF approach relies on this length-scale being smaller than the PDF field of view.  
The later is defined as the length-scale over which the intensity of the experimental PDF signal decays to zero due to finite instrumental resolution width in $\mathbf{Q}$-space. 
This can be established by examining full 3D-PDF calculated in Fourier transformation over a wide range. 
For \oldrobadd{the} \cis\ data discussed here such a view of full 3D-PDF is shown in Fig.~\ref{fig:corr_length}(b).
The intensity scale is selected to bring the strongest features intentionally to saturation in order to expose weaker intensities at large interatomic distance.
Fig.~\ref{fig:corr_length}(b) reveals that the intensity decays to zero by $\sim$35~\AA. 
In contrast, intensity of 3D-$\Delta$PDF (Fig.~\ref{fig:corr_length}(a)) decays to zero by $\sim$10~\AA\ which is well within the field of view of the measurement.}

\esbaddold{\oldrobadd{The observations of }this experiment confirm that local distortions first observed in \oldrobadd{\soutoldold{powder}1D-}PDF~\cite{bozin_local_2019} are also present in \oldrobadd{\soutoldold{crystalline} the single crystal} \cis\ system.
Thorough \rjkadd{\soutold{assessment}characterization} of these observations and their implications for \cis\ are beyond the scope of this report. 
}

\section{Conclusions}
\label{section:conc}
We have extensively outlined the data processing steps necessary for obtaining a \rjkadd{\soutold{reliable}reproducible} 3D-$\Delta$PDF from a single crystal sample, including measurement, crystal orientation, transforming from detector to reciprocal space, handling detector artifacts, merging different data-sets, applying symmetry operations, removing Bragg intensity, and applying a discrete Fourier transform.
These steps have thus far been spread across various different sources, or maintained only as an in-house procedure.

Further, we have investigated aspects of 3D-$\Delta$PDF reproducibility under a number of conditions, including a variation of work-flow ordering, sample to sample variation, and worst case measurement and data processing conditions.
We have found that across \rjkadd{\soutold{all}the majority of} tested situations, the 3D-$\Delta$PDF is remarkably robust, with \rjkadd{\soutold{no}few} qualitative differences.
\rjkadd{Failure to obtain a complete and regularly shaped data-set in reciprocal space, due to a limited measurement range or a low symmetry crystal (or both) can result in spurious features in the 3D-$\Delta$PDF.}
\rjkadd{The 3D-$\Delta$PDF also showed nearly no quantitative differences across the majority of tested situations.
An important exception is the impact of the Bragg punch size.
If the punch is too large, it can affect the decay rate of features in the 3D-$\Delta$PDF.
This is particularly important if the diffuse features of interest are relatively sharp and co-located with Bragg peaks.}
\rjkadd{The 3D-$\Delta$PDF then appears largely robust} even given \rjkadd{a relatively large crystal size, the presence of small polycrystalline inclusions, and the use of a detector with} \esbaddold{\soutoldold{low}limited} dynamic range.
This observation re-enforces the power of this emerging technique, and should underscore observations arising from past and future 3D-$\Delta$PDF studies.

%
%
\ack{
Work at Brookhaven National Laboratory was supported by U.S. Department of Energy, Office of Science, Office of Basic Energy Sciences (DOE-BES) under contract No. DE-SC0012704. 
\rjkadd{We acknowledge DESY (Hamburg, Germany), a member of the Helmholtz Association HGF, for the provision of experimental facilities. Parts of this research were carried out at beamline P21.1 at PETRA III.}
ESB acknowledges the Stephenson Distinguished Visitor Programme for supporting his stay at DESY in Hamburg.  
This research was supported in part through the Maxwell computational resources operated at Deutsches Elektronen-Synchrotron DESY, Hamburg, Germany.
This research was supported in part by the Villum Foundation.
We gratefully acknowledge Olof Gutowski for assistance with the measurements.}

\bibliography{20rjk_3ddpdf_cis}
\bibliographystyle{iucr}
\vfill\newpage

\renewcommand\thefigure{S\arabic{figure}}
\setcounter{figure}{0}



\end{document}